\documentclass[%
 reprint,
 amsmath,amssymb,
 aps,
 prapplied, 
]{revtex4-2}

\usepackage{graphicx}  
\usepackage{dcolumn}   
\usepackage{bm}        
\usepackage{hyperref}  
\usepackage{longtable}
\usepackage{booktabs}
\usepackage[table]{xcolor}
\usepackage{xcolor}
\usepackage{float}
\hypersetup{
    colorlinks = true,
    linkcolor = blue,
    citecolor = blue,
    urlcolor = blue
}
\DeclareUnicodeCharacter{2212}{\ensuremath{-}}
\begin{document}

\title{Phonon-Coupled Hole-Spin Qubits in High-Purity Germanium: Design and Modeling of a Scalable Architecture}

\author{D.-M. Mei}
\author{S. A. Panamaldeniya}
\author{K. Dong}
\author{S. Bhattarai}
\author{N. Budhathoki}
\author{A. Warren}
\affiliation{Department of Physics, University of South Dakota, Vermillion, SD 57069, USA}

\date{\today}

\begin{abstract}

We present a design and modeling of a scalable quantum processor architecture utilizing hole-spin qubits defined in gate-controlled germanium (Ge) quantum dots, where coherent spin-phonon coupling is predicted to facilitate qubit manipulation and long-range interactions. The architecture exploits the strong, electrically tunable spin-orbit interactions intrinsic to hole states in Ge, integrated with high-quality phononic crystal cavities (PnCCs) to enable fully electrical qubit control and phonon-mediated coupling. Employing a streamlined simulation framework built upon multiband \(\mathbf{k}\cdot\mathbf{p}\) modeling and finite-element methods, we quantify key performance metrics, including electrically tunable \( g \)-factors ranging from \(1.3\) to \(2.0\), spin-phonon coupling strengths up to \(6.3\,\mathrm{MHz}\), phononic cavity quality factors exceeding \(10^4\), and phonon-mediated spin relaxation times (\(T_1\)) reaching milliseconds. The proposed architecture concurrently achieves extended spin coherence and rapid gate operations through strategic electric field modulation and engineered phononic bandgap environments. Furthermore, isotopically enriched, high-purity Ge crystals significantly enhance device coherence by minimizing disorder and hyperfine interactions. This integrated approach, merging advanced materials engineering, precise spin-orbit coupling, and phononic cavity design, establishes a promising CMOS-compatible pathway toward scalable, high-fidelity quantum computing.

\end{abstract}

\maketitle

\section{Introduction}

Quantum computing offers a transformative approach to solving computational problems that are intractable for classical systems, such as prime factorization, quantum many-body simulations, and combinatorial optimization. Central to these capabilities is the quantum bit, or qubit, which can exist in superpositions and entangled states—enabling massively parallel computation. Over the past two decades, various physical implementations of qubits have been explored, each presenting distinct advantages and limitations in terms of coherence, control, and scalability.

Among the leading platforms are superconducting qubits~\cite{Kjaergaard2020}, trapped ions~\cite{Blatt2012}, and semiconductor-based quantum dots~\cite{Veldhorst2015}. Superconducting circuits offer fast gate operations and compatibility with microwave control infrastructure but face coherence limitations and challenges with scaling due to crosstalk and frequency crowding. Trapped ions provide exceptional coherence and high-fidelity gates, though their reliance on complex laser-based control architectures and slower gate speeds pose practical constraints for large-scale integration.

Semiconductor spin qubits—particularly those realized in quantum dots—represent a compelling pathway toward scalable, electrically controlled quantum processors. Early demonstrations in GaAs showed coherent manipulation of single electron spins~\cite{Petta2005}, but decoherence from nuclear spin environments limited performance. This limitation has prompted a shift to group-IV semiconductors such as silicon (Si) and germanium (Ge), where isotopic purification and enrichment can nearly eliminate nuclear spin noise.

Si spin qubits have achieved long coherence times~\cite{Veldhorst2014} and have been implemented using industry-standard fabrication techniques~\cite{Zwerver2022}. However, Si’s inherently weak spin-orbit coupling restricts electrical tunability and gate speed. In contrast, Ge offers a superior materials platform: it supports strong and tunable spin-orbit interaction~\cite{Watzinger2018}, exhibits high hole mobilities in two-dimensional heterostructures, and can be isotopically purified and enriched to suppress decoherence from nuclear spins. These characteristics enable fast, all-electrical spin control within gate-defined quantum dot architectures.

Recent experiments with strained Ge/SiGe heterostructures have demonstrated robust confinement of hole spins, long coherence times, and precise electrical tunability~\cite{Hendrickx2020}. Ge hole-spin qubits are emerging as promising candidates for scalable, high-fidelity quantum operations. A comprehensive roadmap for Ge-based quantum electronics by Scappucci \textit{et al.}~\cite{Scappucci2020} highlights its integration compatibility with planar architectures and the utility of strain engineering for optimizing quantum dot properties.

While phonons—quantized lattice vibrations—are often viewed as sources of decoherence, recent theoretical and experimental studies suggest they can instead be engineered as active resources for qubit control. In materials with strong spin-orbit coupling (SOC), such as Ge, phonon-mediated transitions enable electric-field-controlled spin flips and phonon-assisted readout or coupling mechanisms~\cite{Maier2013, Bosco2021}. This opens the door to hybrid spin-phonon architectures with electrically tunable interactions.

Phononic crystal cavities (PnCCs), which confine GHz-frequency acoustic modes with high quality factors, provide the analog of optical cavities in cavity quantum electrodynamics (QED). When co-integrated with Ge quantum dots, these cavities can serve as mediators for long-range qubit-qubit coupling, boost spin readout fidelity via phonon emission detection, or stabilize spin states through phonon filtering. Such integration supports the development of scalable solid-state quantum systems that combine coherence, speed, and electrical control.

Experimental evidence supports the feasibility of this approach. Phonon confinement has been successfully demonstrated through phononic cavity engineering in heterostructures~\cite{Katsaros2011}, enabling both suppression of unwanted decoherence channels and precise control of phonon mode dispersion. These capabilities are essential for enhancing spin-phonon interaction strength—an enabling mechanism for qubit control, readout, and entanglement in Ge-based systems.

\begin{figure}[H]
  \centering
  \includegraphics[width=0.45\textwidth]{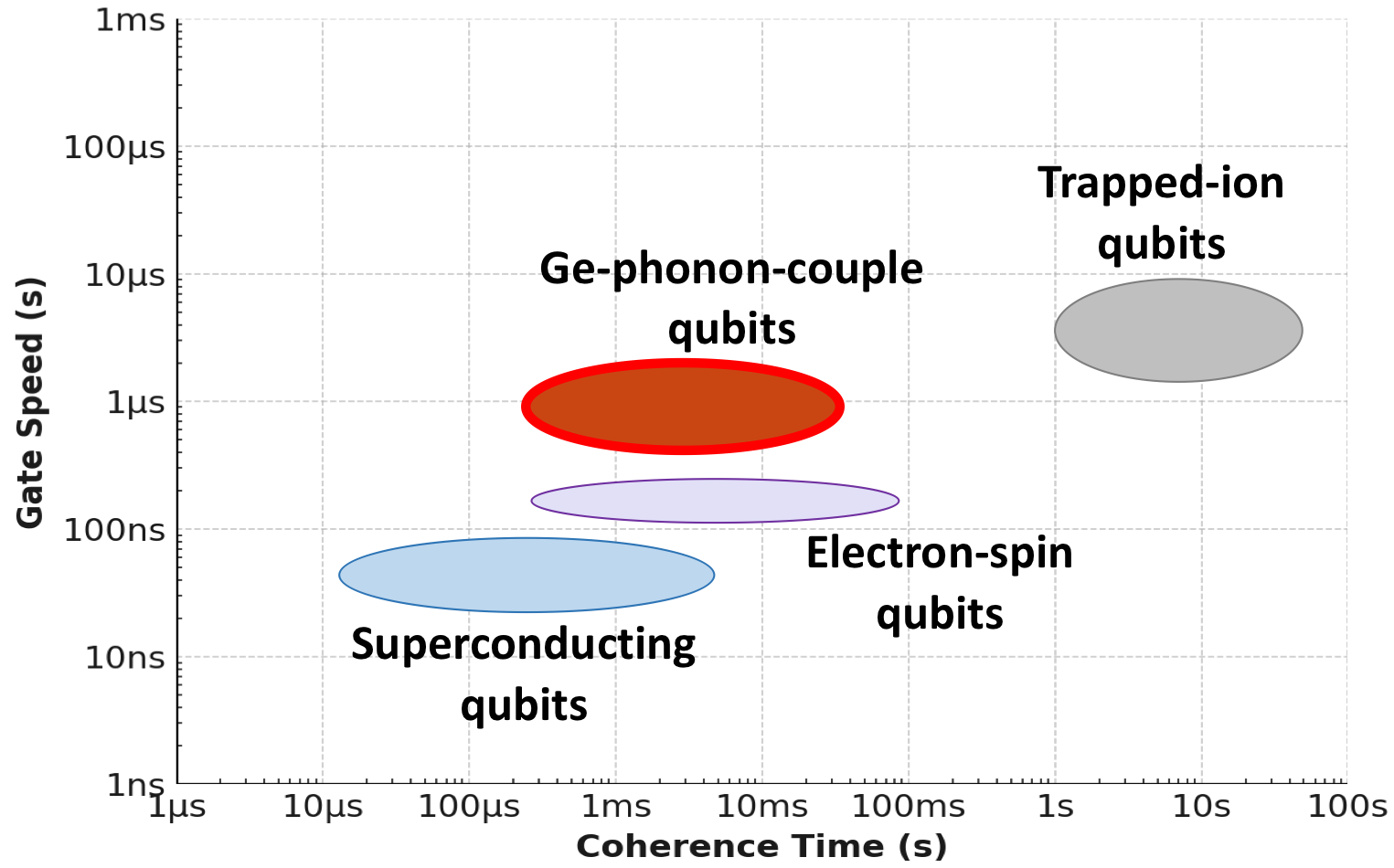}
  \caption{Comparison of representative qubit technologies based on coherence time and gate speed, reflecting current experimental capabilities~\cite{Kjaergaard2020, Blatt2012, Hendrickx2020, Scappucci2020}. Ge-based hole-spin qubits with phonon coupling are positioned in a favorable regime that combines fast, all-electrical control with long coherence, underscoring their promise for high-fidelity quantum information processing. Ge-phonon-coupled reflects \emph{two-qubit} rates mediated by a phonon bus (dispersive exchange), typically \(\sim 0.5\text{--}1~\mu\mathrm{s}\), whereas \emph{single-qubit} in Ge remains \(\mathcal{O}(10\text{--}50\,\mathrm{ns})\) under our operating conditions (see Sec.\,IV).}
  \label{fig:qubitlandscape}
\end{figure}

To contextualize the proposed Ge-based phonon-coupled hole-spin qubit within the broader quantum computing landscape, Fig.~\ref{fig:qubitlandscape} presents a comparative overview of several leading qubit modalities, using coherence time and gate speed as primary performance benchmarks. The figure highlights where current experimental realizations stand with respect to these two key metrics.

Within this landscape, we benchmark Ge hole--spin qubits against leading alternatives beyond superconducting circuits and trapped ions, namely \emph{neutral atoms} in optical tweezers (including Rydberg--mediated entangling gates) and \emph{solid--state defect centers} (e.g., NV centers in diamond). These modalities achieve competitive coherence and high--fidelity control but impose distinct hardware and integration constraints (optical access, cryogenic operation, and packaging), which set their operating points in the gate--speed/coherence plane. To make the comparison concrete, Fig.~\ref{fig:qubitlandscape} includes representative single-- and two--qubit time scales and coherence values for these platforms; the caption reports typical experimental ranges to provide context for the Ge results highlighted here.

While scalability is not depicted in Fig.~\ref{fig:qubitlandscape}, it is an equally important consideration in evaluating qubit platforms. In this work, scalability is treated as a qualitative metric that reflects the practical feasibility of large-scale quantum processor integration. This includes factors such as compatibility with planar CMOS fabrication, support for densely packed qubit arrays, use of purely electrical control mechanisms, and architectural suitability for parallel operations and quantum error correction. The discussion in the main text draws upon established criteria, particularly those outlined by Scappucci et al.~\cite{Scappucci2020}, to assess the integration-readiness of various platforms.

Among the technologies compared, Ge-based hole-spin qubits with phonon coupling offer a unique combination of millisecond-scale coherence times and nanosecond-scale gate speeds, a regime that enables fast, coherent control while maintaining compatibility with scalable semiconductor manufacturing. In contrast to platforms such as trapped ions or superconducting qubits, which face inherent constraints from optical systems or complex cryogenic wiring, the proposed Ge platform leverages phonon-engineered spin-orbit interactions and mature fabrication techniques to simultaneously optimize coherence, control speed, and integration potential.

A critical factor influencing the performance of Ge-based hole-spin qubits is material quality, particularly impurity scattering and charge noise, which significantly degrade qubit coherence. To address these challenges, this study leverages in-house capabilities at the University of South Dakota (USD) for growing detector-grade, high-purity Ge crystals, achieving impurity concentrations as low as \(\sim 10^{10}\,\text{cm}^{-3}\)~\cite{gang1,gang2,gang3,wang1,wang2,wang3,sanjay1}. This vertically integrated crystal growth infrastructure provides a robust pathway for fabricating low-disorder quantum dots, thus enabling detailed exploration of phonon-assisted spin manipulation within exceptionally clean Ge-based systems.

A distinctive strength of our proposed architecture lies in the strategic integration of isotopically enriched, high-purity \(^{74}\)Ge substrates, the strong and electrically tunable spin-orbit coupling inherent to hole states, and precisely engineered phononic crystals (PnCs) with embedded PnCCs. Together, these components form a robust platform for spin-phonon-based quantum information processing. The PnC structure is designed to surround the quantum dot, creating a phonon bandgap that suppresses unwanted phonon modes and isolates a target frequency range (e.g., around 6\,GHz). Localized defects introduced into the PnC form PnCCs that support discrete acoustic modes resonant with the desired frequency. These modes are spatially aligned with the wavefunction of the hole-spin qubit in the Ge quantum dot, enabling coherent spin-phonon interaction via the deformation potential. \textit{Related theory for Ge spins embedded in quasi-2D phononic crystals shows that tuning the Zeeman splitting into a phononic bandgap suppresses one-phonon relaxation and enables long-range coupling via virtual phonon exchange, which we extend here to Ge hole-spin dots}~\cite{Smelyanskiy2014a}. While the PnC acts as an acoustic shield to minimize phonon leakage and external noise, the PnCC enhances the local phonon density of states, thereby increasing the coupling strength. This integrated approach facilitates strong, localized spin-phonon coupling essential for coherent qubit control, high-fidelity readout, and phonon-mediated qubit entanglement.

Unlike prior studies that have investigated spin-orbit effects or phononic engineering in isolation, our architecture synergistically merges these two domains to realize all-electrical spin manipulation along with phonon-mediated qubit coupling. The use of isotopically purified \(^{74}\)Ge substrates significantly reduces hyperfine-induced decoherence, resulting in extended qubit coherence times. The integrated PnC and PnCC structures act as both spectral filters and high-Q resonators, selectively enhancing desired phonon-mediated processes while suppressing incoherent environmental interactions. Collectively, this materials–device integration strategy positions Ge-based hole-spin qubits as highly promising candidates for scalable, high-fidelity quantum computing platforms.

Beyond its conceptual and theoretical strengths, our proposed architecture also accounts for practical implementation challenges, such as surface roughness, defect formation during lithographic processing, and strain variability arising from lattice mismatch. These issues can be effectively mitigated through a combination of optimized dry and wet etching processes, surface passivation and annealing protocols, and the use of lattice-matched Si$_{1-x}$Ge$_x$ buffer layers or isotopically enriched \(^{74}\)Ge substrates. By incorporating these fabrication strategies, the architecture enhances the feasibility of experimentally realizing phonon-coupled qubit devices with extended coherence times, thereby bridging the gap between simulation and scalable quantum device implementation.

In this work, we model and evaluate an integrated device platform that combines gate-defined hole-spin qubits in isotopically enriched Ge quantum dots with engineered PnC and PnCCs. Leveraging a streamlined simulation framework that integrates multiband \textit{k·p} modeling with finite-element-based calculations of strain and electric field profiles, we systematically analyze electrically tunable \(g\)-factors, spin-phonon coupling strengths, spin relaxation times (\(T_1\)), and phononic cavity quality factors (\(Q\)). Our findings demonstrate that phonon-assisted qubit manipulation and coupling are not only physically realizable but also highly tunable via external gating and nanoscale structural design. These results provide a solid foundation for building scalable, electrically controlled solid-state quantum processors that employ phonons as engineered resources for quantum information science.

\section{Spin-Phonon Coupling in Germanium Quantum Dots}

A central feature of phonon-assisted qubit control in Ge quantum dots is the spin-phonon interaction, which enables transitions between spin states via lattice vibrations. This mechanism, rooted in strong SOC and valence-band structure, is particularly advantageous in Ge due to its high intrinsic SOC and compatibility with electrostatic control. Understanding this interaction provides the foundation for designing qubit architectures that leverage phonons not merely as environmental noise sources, but as engineered resources for initialization, control, and coupling.

\begin{figure}[H]
  \centering
  \includegraphics[width=0.45\textwidth]{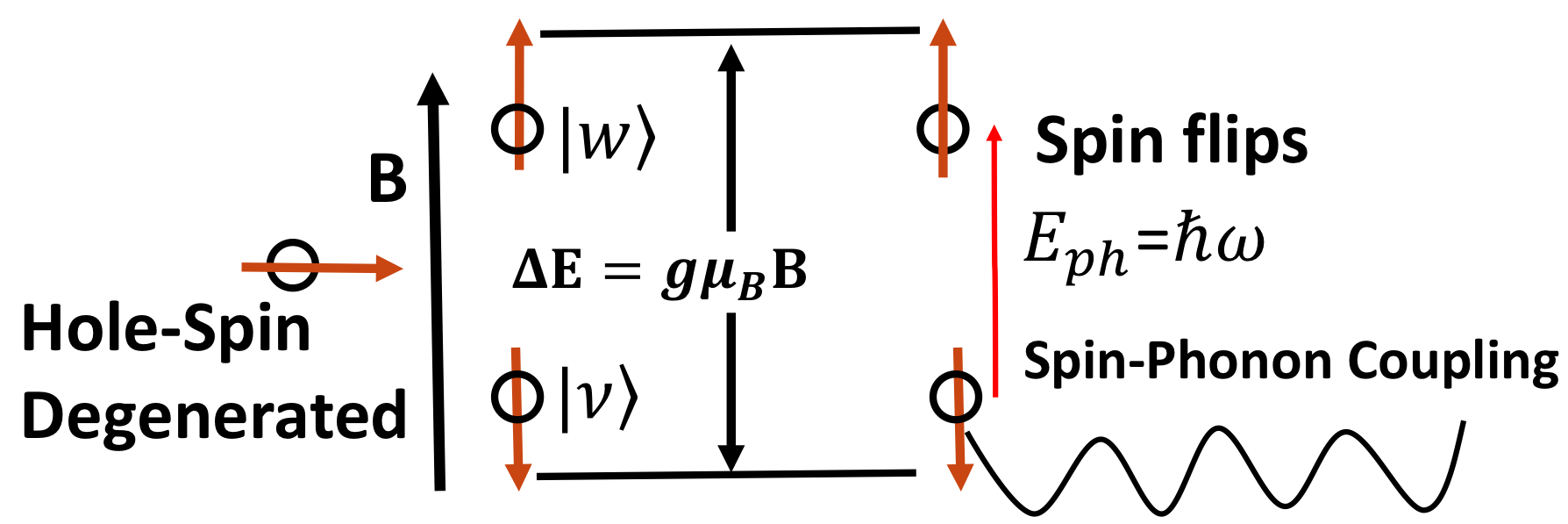}
  \caption{Schematic showing spin-phonon coupling in a Ge quantum dot.  \textbf{Left:} In zero magnetic field, the hole-spin states are degenerate. \textbf{Middle:} With an external magnetic field applied along [001], the states split by the Zeeman energy $\Delta E=g\mu_B B$; phonons with energy $E_{\rm ph}=\hbar\omega=\Delta E$ can induce spin-flip transitions via spin--orbit coupling, which are enhanced by heavy-hole/light-hole mixing in low-symmetry environments. Here, $\lvert v\rangle$ and $\lvert w\rangle$ denote the spin-down and spin-up eigenstates of the spin qubit in the Ge quantum dot. \textbf{Right:} A hole initially in $\lvert v\rangle$ (\emph{spin-down}) can absorb a matching-energy phonon and transition to $\lvert w\rangle$ (\emph{spin-up}).}

  \label{fig:spinphonon}
\end{figure}

Figure~\ref{fig:spinphonon} schematically illustrates the spin-phonon coupling mechanism in a Ge quantum dot subjected to an external magnetic field. The field induces Zeeman splitting between hole-spin states, given by~\cite{Winkler2003, Maier2013}:
\begin{equation}
\Delta E = g \mu_B B,
\label{eq:zeeman}
\end{equation}
where \( g \) is the effective Landé g-factor and \( \mu_B \) is the Bohr magneton. In Ge, the g-factor is both anisotropic and tunable via gate-defined electric fields and local strain, enabling electrically mediated spin-state control~\cite{Kloeffel2018, Marcellina2020}. When a lattice vibration (phonon) with energy \( E_{ph} = \hbar \omega \) matches the Zeeman energy splitting, it can mediate spin-flip transitions between these states.

\noindent

While Eq.~\eqref{eq:zeeman} assumes an isotropic $g$-factor for simplicity, we note that in strained Ge quantum wells, the $g$-factor and spin–phonon coupling are in fact anisotropic~\cite{Maier2013, Terrazos2021}. This anisotropy arises from the complex valence band structure and spin–orbit interactions, which are sensitive to the orientation of the applied magnetic and electric fields relative to the crystallographic axes. Specifically, both the magnitude and sign of the effective $g$-factor can vary with the polar and azimuthal angles of the field vectors~\cite{Kloeffel2011}. As such, optimal spin–phonon coupling is generally achieved when the magnetic field is aligned along the growth direction (typically [001]) and the electric field lies in the quantum well plane. Misalignment can lead to reduced coupling efficiency or modified spin splitting, which may affect gate fidelity and readout. In future work, the angular dependence of $g_{\mathrm{sp}}$ and its implications for qubit control fidelity will be modeled more explicitly.

The effectiveness of spin-phonon coupling depends on several key factors. First, the strength of SOC in Ge is significantly enhanced for holes due to their p-orbital character. Second, the confinement potential symmetry in the quantum dot determines the degree of heavy-hole and light-hole mixing, which facilitates phonon-induced transitions~\cite{Maier2013}. Third, the phonon mode characteristics—such as frequency, polarization, and spatial localization—govern how efficiently phonons can couple to spin states.

At cryogenic temperatures (\(<1~\text{K}\)) relevant for quantum computing, acoustic phonons dominate over optical modes. The primary spin relaxation mechanism in this regime is acoustic deformation potential coupling, whereby phonons modulate the local band structure and mediate transitions between spin states~\cite{Watzinger2018, Scappucci2020}. This spin-phonon interaction plays a dual role: while it enables efficient spin initialization and readout via phonon emission or absorption, it can also introduce decoherence through relaxation ($T_1$) and dephasing ($T_2^*$) pathways.

To mitigate these effects and simultaneously enable coherent control, it is advantageous to operate the qubit at frequencies compatible with standard microwave electronics and engineered phononic environments. In particular, targeting the 2--6~GHz frequency range allows for integration with superconducting resonators and phononic cavities designed for enhanced spin-phonon coupling. To achieve spin-flip transitions in this range, the Zeeman splitting must fall within approximately 8 to 25~$\mu$eV.

\begin{figure}[htp]
    \centering
    \includegraphics[width=0.45\textwidth]{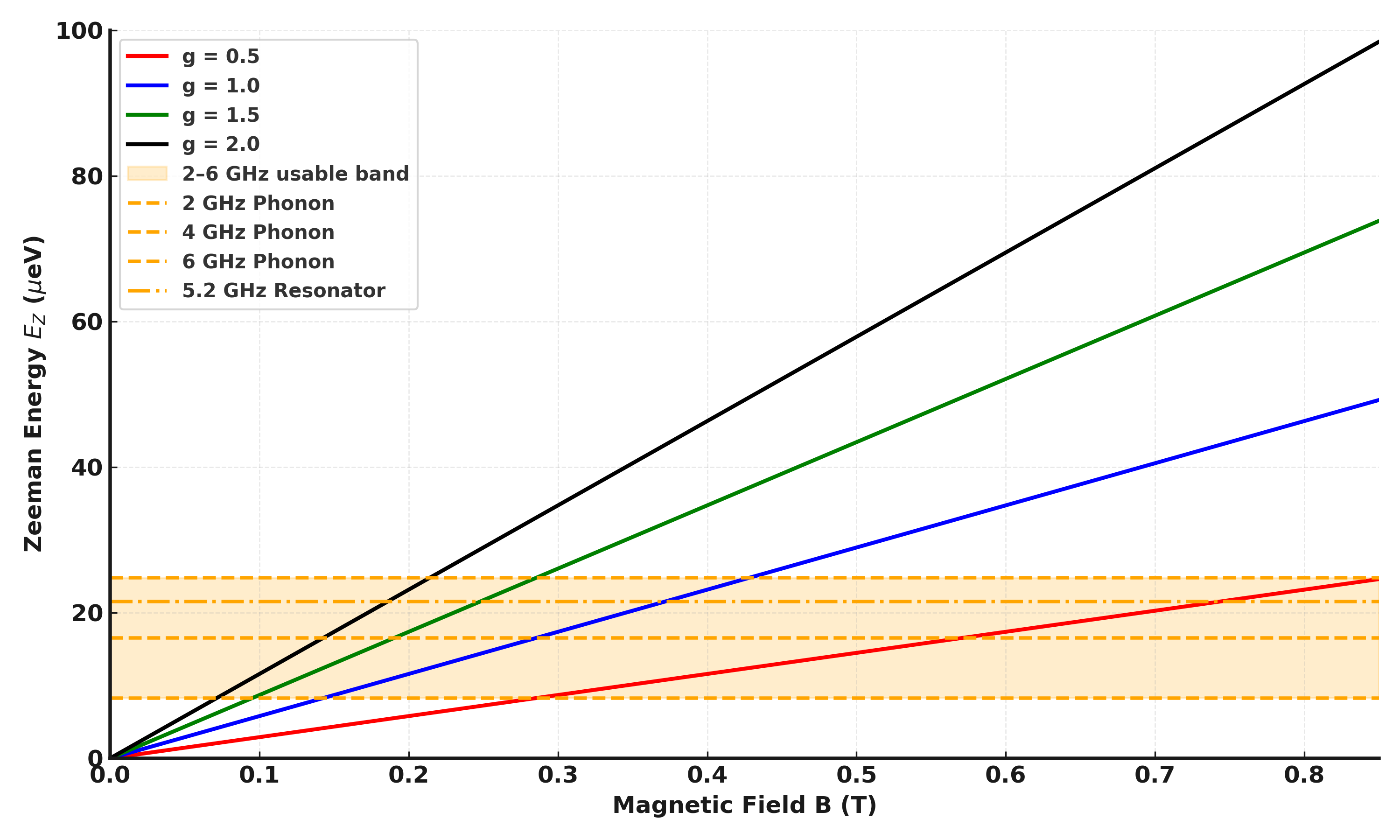} 
    \caption{
        Zeeman energy as a function of magnetic field $B$ for various hole $g$-factors, overlaid with phonon and resonator mode energies. 
        Each curve shows the Zeeman splitting $\Delta_E = g \mu_B B$ for hole spin qubits with tunable $g$-factors in the range $g = 0.5$ to $2.0$. 
        The shaded region indicates the usable microwave control band (2--6 GHz), corresponding to Zeeman energies between approximately 8 and 25~$\mu$eV. 
        Horizontal dashed lines represent phonon mode energies at 2, 4, and 6~GHz, and the dash-dotted red line marks a typical superconducting resonator frequency at 5.2~GHz. 
        }
    \label{fig:zeeman_vs_B}
\end{figure}

As illustrated in Figure~\ref{fig:zeeman_vs_B}, this condition can be met by tuning the hole $g$-factor and applied magnetic field. For example, taking an \emph{in-plane, electrically tunable} effective $g$-factor $g_{\rm eff}\approx1.3$---consistent with strained-Ge quantum wells and dots under finite vertical electric field---a field of $B=0.25$~T yields a Zeeman energy of roughly 19~$\mu$eV, enabling microwave-frequency control~\cite{Terrazos2021, Marcellina2020, Sarkar2023, Scappucci2020}. This contrasts with out-of-plane $g$-factors $g_z\gtrsim5$--$8$ reported for perpendicular fields and different device geometries.

Operating in this energy range also imposes constraints on the thermal environment: to suppress thermal activation and ensure reliable spin-state initialization, temperatures below 100~mK are typically required.
From a device engineering standpoint, the dual nature of spin-phonon interactions can be harnessed by shaping the phonon density of states. Techniques such as phononic bandgap engineering and the incorporation of resonant acoustic cavities can be employed to suppress unwanted relaxation channels while enhancing desired transitions~\cite{Niquet2022}. 
Moreover, recent theoretical studies have shown that spin-phonon coupling in Ge can be made anisotropic and tunable through device geometry and electric fields~\cite{Bosco2021}. This tunability enables all-electrical control of spin states without relying on microwave magnetic fields, offering a scalable path toward high-fidelity quantum operations. These capabilities highlight the unique advantages of Ge-based platforms for developing hybrid spin-phonon quantum architectures.

\section{Phonon-Assisted \( T_1 \) Relaxation}

In addition to enabling all-electrical spin control, spin-phonon coupling plays a key role in limiting qubit performance by contributing to spin relaxation. Specifically, the longitudinal relaxation time \( T_1 \), which characterizes the decay of an excited spin state to its ground state, is predominantly governed by phonon-mediated processes in systems with strong SOC. Understanding and engineering this relaxation mechanism is critical for achieving long-lived, high-fidelity qubits in Ge quantum dot platforms.

\begin{figure}[H]
  \centering
  \includegraphics[width=0.5\textwidth]{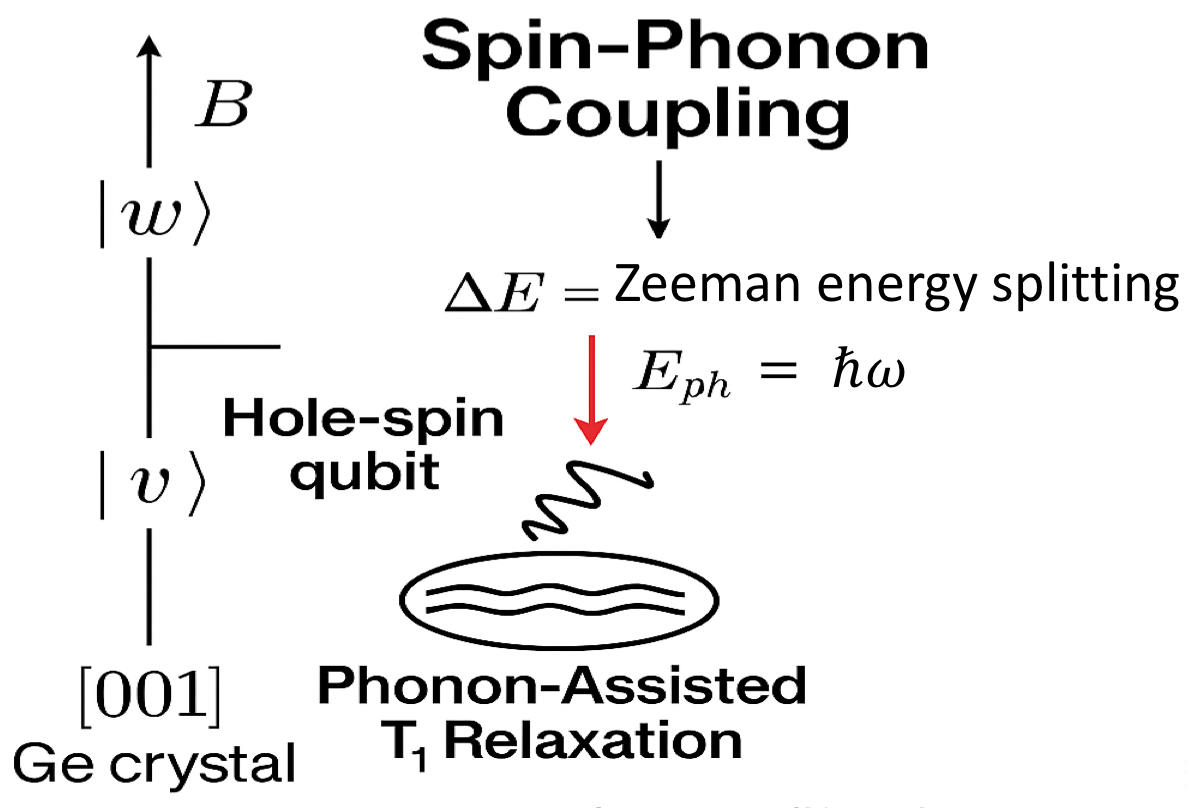}
  \caption{Phonon-assisted relaxation process in a Ge qubit. \textbf{Left:} In an external magnetic field, the excited spin state relaxes to the ground state via spontaneous emission of a phonon with energy \(E_{\rm ph}=\hbar\omega=\Delta E\) (\textbf{center}).
This one-phonon process is enabled by spin--orbit coupling and is influenced by heavy-hole/light-hole mixing as well as device-specific factors—quantum-dot geometry, crystal orientation, and the phonon-mode structure (bulk substrate, phononic crystal, or cavity).}
  \label{fig:relaxation}
\end{figure}

As illustrated in Fig.~\ref{fig:relaxation}, a hole-spin qubit in the excited state can spontaneously emit a phonon and transition to the ground state when the emitted phonon’s energy matches the Zeeman splitting described in Eq.~(1). This process is made possible by SOC, which mixes spin and orbital degrees of freedom, allowing spin-flip transitions via lattice strain-induced perturbations.

The spin relaxation rate depends on both the strength of the spin-phonon interaction and the phonon density of states at the qubit transition energy. Quantitatively, the relaxation rate is expressed as:
\begin{equation}
\frac{1}{T_1} = \frac{2\pi}{\hbar} |M_{\text{sp}}|^2 D(\omega),
\label{eq:msp}
\end{equation}
where \( M_{\text{sp}} \) is the spin-phonon matrix element and \( D(\omega) \) is the phonon density of states~\cite{Maier2013, Bosco2021}. In Ge quantum dots, this coupling arises primarily from acoustic deformation potential interactions and Rashba-type SOC, both of which are tunable through gate-controlled electric fields~\cite{Maier2013, Marcellina2020}.

At millikelvin temperatures, the phonon bath is dominated by low-energy acoustic modes, which typically reduce the relaxation rate. Moreover, when the phonon wavelength exceeds the quantum dot size, relaxation can be strongly suppressed due to the \textit{phonon bottleneck effect}~\cite{Khaetskii2001, Trif2009}. This arises from poor overlap between the qubit’s spatial wavefunction and long-wavelength phonon modes, reducing the efficiency of phonon emission.

The heterostructure design—including quantum dot geometry, crystal orientation, and strain distribution—plays a pivotal role in determining spin relaxation dynamics. Strain-induced mixing between heavy-hole and light-hole states can significantly enhance spin-orbit coupling (SOC), increasing the spin-phonon interaction matrix element \( M_{\text{sp}} \) and accelerating relaxation processes~\cite{Bosco2021, Wang2021}. Additionally, the orientation of the applied magnetic field relative to the crystallographic axes introduces anisotropy in the spin-phonon coupling, offering a valuable tuning parameter for optimizing qubit performance.

Quantitatively, longitudinal relaxation times \( T_1 \) in Ge hole-spin qubits can span from microseconds to milliseconds, depending on key factors such as confinement geometry, electric field strength, and magnetic field orientation~\cite{Hendrickx2020, Jirovec2021}. For example, increasing the vertical electric field strengthens Rashba-type SOC~\cite{Watzinger2018, Marcellina2020}, which enhances spin control fidelity but also increases the spin relaxation rate~\cite{Maier2013}. Strong vertical confinement in narrow Ge quantum wells increases orbital level separation, suppressing orbital mixing and thereby reducing spin-phonon coupling~\cite{Kloeffel2018}. Conversely, lateral confinement asymmetry can amplify heavy-hole/light-hole mixing, enhancing \( |M_{\text{sp}}| \) and accelerating relaxation~\cite{Wang2021}. Experimentally, higher magnetic fields have been shown to shorten \( T_1 \) by expanding the phonon phase space available at increased Zeeman energies~\cite{Hendrickx2020, Bosco2021}. These dependencies highlight the importance of careful tuning of confinement, gating, and magnetic field orientation. When integrated with phononic engineering techniques such as cavity-induced spectral shaping~\cite{Scarlino2022, Midolo2018}, this design space enables precise control over \( T_1 \), allowing system-level optimization for both coherence and controllability.

To suppress undesired relaxation pathways while retaining phonon-mediated functionality, the phononic environment can be precisely engineered using structures such as PnCCs with acoustic bandgap materials (PnCs). These elements reshape the phonon density of states, selectively suppressing environmental decoherence channels while enhancing coupling to desirable vibrational modes. By confining phonons to discrete, high-quality-factor modes resonant with the qubit transition frequency, PnCCs can significantly extend spin lifetimes, enable high-fidelity dispersive readout, and facilitate phonon-mediated coupling between spatially separated qubits.

\begin{figure}[htp]
  \centering
  \includegraphics[width=0.45\textwidth]{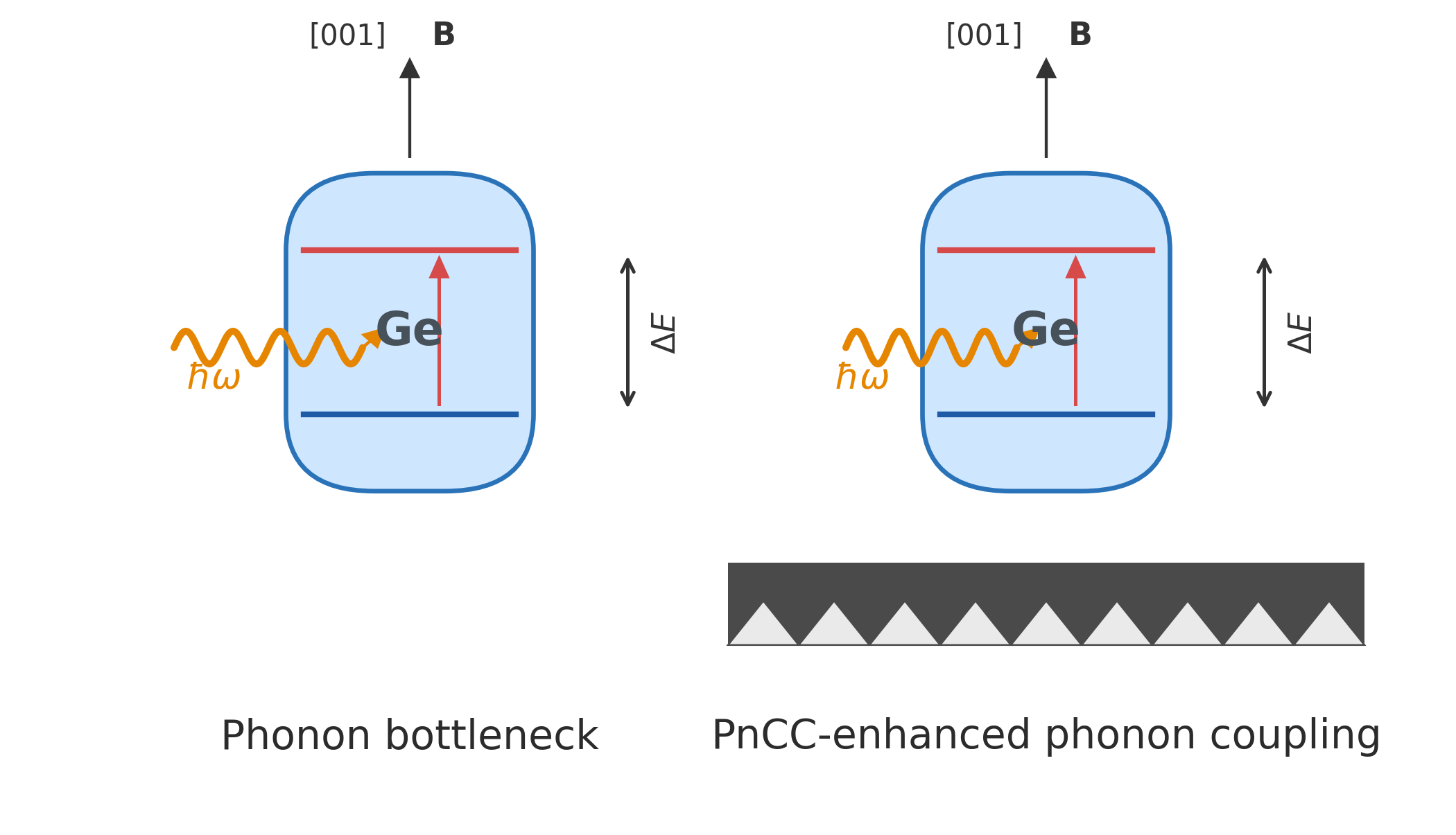} 
  \caption{Comparison of phonon bottleneck and cavity-enhanced relaxation regimes.  \textbf{Left:} In small quantum dots, when the dominant phonon wavelength \(\lambda_{\rm ph}\) exceeds the dot size \(L\), the spin--phonon matrix element is suppressed by the dot form factor (the “phonon bottleneck”), yielding longer \(T_1\). This panel is the reference (no phononic crystal).
\textbf{Right:} A 2D phononic-crystal cavity (PnCC) creates a localized defect mode inside an engineered bandgap, concentrating the phonon density of states at the qubit transition \((\hbar\omega_Z=\Delta E)\) and thereby enhancing spin--phonon coupling; off resonance (within the bandgap) relaxation remains suppressed, while on resonance with the defect mode it can be intentionally increased or used for engineered interactions. The diagonally hatched gray band marks the phononic bandgap where propagating acoustic modes are suppressed, reducing \(T_1^{-1}\). The color scale and axes are identical in both panels for direct comparison.}
  \label{fig:phononbottleneck}
\end{figure}

Figure~\ref{fig:phononbottleneck} contrasts these two regimes. In the phonon bottleneck scenario (left), relaxation is inhibited due to a mismatch between the phonon wavelength and quantum dot dimensions. In contrast, a properly designed PnCC (right) enhances relaxation or control by concentrating phonon density at the desired transition frequency.

Although this work primarily focuses on spin relaxation processes characterized by the longitudinal relaxation time \( T_1 \), dephasing mechanisms quantified by the inhomogeneous coherence time \( T_2^* \) are equally critical for determining the overall fidelity of spin--phonon operations. In Ge-based hole-spin qubit systems, \( T_2^* \) is typically limited by low-frequency charge noise and slow drift in the electrostatic confinement potential~\cite{Yoneda2018, Scarlino2022}, rather than by hyperfine interactions, which are strongly suppressed in isotopically enriched \(^{74}\)Ge~\cite{Scappucci2020, Jirovec2021}.

Importantly, the use of PnCCs in our architecture offers a powerful mechanism to enhance \( T_2^* \) by mitigating spectral diffusion caused by environmental phonons. The phononic bandgap suppresses broadband acoustic noise, while high-Q confinement restricts phonon-mediated coupling to discrete, well-defined vibrational modes. This engineered suppression of the phonon density of states effectively isolates the qubit from low-frequency strain fluctuations and acoustic background, thereby preserving phase coherence of the spin state.

Under optimized gating and environmental control, we estimate that \( T_2^* \) in this architecture could exceed \( 1\text{--}10~\mu\text{s} \), consistent with or exceeding values reported in other planar Ge and Si-based platforms~\cite{Hendrickx2020, Jirovec2021}. Such coherence times are compatible with the simulated spin relaxation time \( T_1 \sim \text{ms} \), and sufficient to support high-fidelity spin readout, control, and phonon-mediated entanglement operations.

Beyond the inhomogeneous dephasing time, the intrinsic coherence time \( T_2 \) provides a more fundamental measure of phase coherence, representing the timescale for irreversible decoherence processes such as spin-bath interactions or high-frequency phonon scattering. \( T_2 \) is typically extracted using spin-echo or dynamical decoupling protocols that refocus slowly varying noise components~\cite{Veldhorst2014, Yoneda2018}. In our proposed design, several features are expected to positively influence \( T_2 \): the use of isotopically enriched Ge eliminates nuclear-spin-induced dephasing; the phononic bandgap suppresses incoherent phonon scattering; and cavity-induced spectral narrowing reduces residual environmental coupling~\cite{ArrangoizArriola2019, Gely2021}. Although detailed modeling of \( T_2 \) is beyond the present scope, we anticipate that echo-based coherence times in the range of \( 50\text{--}100~\mu\text{s} \) are achievable, consistent with recent experimental results~\cite{Jirovec2021, Huang2019, Muhonen2014}. Furthermore, the incorporation of charge-noise-insensitive “sweet spots”~\cite{Wang2021} and advanced decoupling sequences may further extend \( T_2 \), positioning this platform for high-fidelity, scalable quantum operations mediated by phonons.

\section{Phononic Crystal Cavity for Control and Coupling}

Harnessing and manipulating phonons for coherent qubit control requires engineering their spatial and spectral properties. PnCCs offer a robust solution by confining acoustic phonons to sub-micron regions and tailoring their frequencies to match specific qubit transitions. These cavities operate by introducing periodic modulations in elastic or geometric properties—such as etched hole arrays or alternating material stacks—to create phononic bandgaps, analogous to photonic bandgaps in cavity QED systems~\cite{Safavi2010, Mohammadi2009}. Localized defect modes embedded within these bandgaps allow discrete phonon modes to be confined with high quality factors (\(Q > 10^4\)), making them ideal mediators of spin-phonon interactions in solid-state qubits.

\begin{figure} [H]
  \centering
  \includegraphics[width=0.45\textwidth]{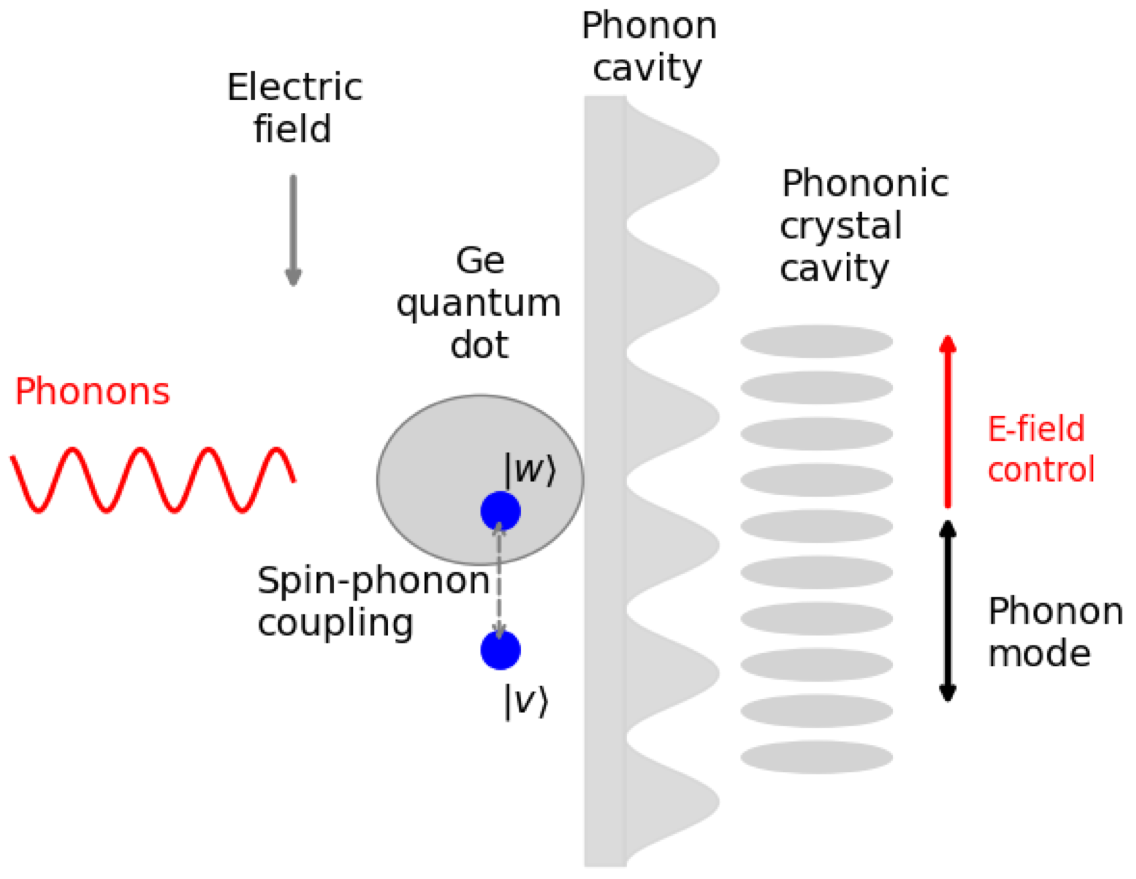} 
  \caption{\emph{Phononic-crystal cavity concept and qubit coupling.} \textbf{Left:} Gate-defined Ge hole-spin quantum dot with adjacent control/readout gates; the dot is positioned near a displacement antinode to maximize the spin–phonon coupling \(g_{\rm sp}\).
\textbf{Middle:} A point defect in a 2D phononic crystal confines a GHz mechanical mode inside the bandgap; the localized strain field overlaps the dot and can be driven by gate modulation, enabling interaction with the spin via spin–orbit coupling.
\textbf{Right:} DC electric-field control (\(E_z\)) tunes the effective \(g\)-factor and Zeeman splitting \(\Delta E=\hbar\omega_Z\), allowing spectral alignment (or deliberate detuning) with the cavity mode for controlled coupling and relaxation.}
  \label{fig:phononcavity}
\end{figure}

The proposed device architecture integrates a gate-defined Ge quantum dot with a nearby or underlying PnCC, as schematically shown in Fig.~\ref{fig:phononcavity}. The quantum dot is formed in a strained Ge quantum well—typically sandwiched between SiGe barriers—using surface gates defined by electron-beam lithography. These gates enable precise control of the dot potential and spin-orbit coupling strength. Beneath the dot, a patterned PnCC confines acoustic phonons in the GHz range, with spectral alignment to the qubit energy splitting \( \Delta E = g \mu_B B \). This alignment permits resonant spin-flip transitions mediated by single-phonon emission or absorption.

The interaction between the qubit and the cavity mode is strengthened by the strong spin--orbit coupling intrinsic to Ge’s valence band, along with the spatial overlap between the phonon mode and the confined hole wavefunction. Real-time modulation of the qubit--cavity detuning is achieved through electric field tuning of the $g$-factor~\cite{Watzinger2018}, providing dynamic control over qubit--phonon interactions. This tunability is crucial for enabling phonon-mediated two-qubit gates, dispersive readout schemes, and coherent state transfer across distant nodes~\cite{Stockill2019, ArrangoizArriola2019, Sletten2019}.

Building on this foundation, we describe how the PnCC architecture facilitates coherent coupling between multiple Ge-based hole-spin qubits to support two-qubit gate operations. For single-qubit control, the platform employs electric dipole spin resonance (EDSR), which harnesses the strong spin--orbit interaction in strained Ge quantum wells. Oscillating gate voltages drive spin rotations, eliminating the need for microwave magnetic fields~\cite{Watzinger2018, Hendrickx2020}. Simulations and prior experimental results demonstrate gate times in the 10--50~ns range, enabling fast, all-electrical qubit manipulation well within the available coherence time.

For two-qubit operations, we consider a phonon-mediated dispersive coupling mechanism wherein two spatially separated hole-spin qubits interact through a shared confined phonon mode localized in the PnCC. This mode serves as a virtual mediator analogous to a cavity photon in circuit QED. When both qubits are coupled to the same phonon mode with coupling strength \(g_{\text{sp}}\) and are operated at a large detuning \(\Delta\) from the phonon resonance, the resulting second-order perturbative interaction yields the effective Hamiltonian:
\begin{equation}
\label{eq:Eq3}
H_{\text{eff}} = \frac{g_{\text{sp}}^2}{\Delta} \left( \sigma_+^{(1)} \sigma_-^{(2)} + \sigma_-^{(1)} \sigma_+^{(2)} \right),
\end{equation}
where \(\sigma_+^{(i)} = |1\rangle_i\langle 0|\) and \(\sigma_-^{(i)} = |0\rangle_i\langle 1|\) are the spin raising and lowering operators for qubit \(i\).  Eq.~\ref{eq:Eq3} follows from the deformation-potential (Bir--Pikus) spin--phonon Hamiltonian after quantizing the elastic field and projecting onto the qubit subspace; see Refs.~\cite{Rabl2010, Bulaev2005, Li2020, Burkard2023} for detailed derivations in related solid-state platforms.

This Hamiltonian represents an effective XY-type interaction, commonly encountered in cavity and circuit QED systems~\cite{Majer2007, Blais2021}. Although it is not a direct exchange interaction---which requires significant overlap of wavefunctions---it produces a functionally similar spin-flip-flop dynamic through coherent virtual phonon exchange. Thus, while it differs physically from exchange coupling, the effective result---entangling operations between spin qubits---is operationally analogous.

With typical parameters of \(g_{\text{sp}}/2\pi \sim 5~\text{MHz}\) and detuning \(\Delta/2\pi \sim 50~\text{MHz}\), the effective two-qubit gate rate lies in the range of 0.25--1~MHz, corresponding to entangling gate durations of approximately 0.5--1~\(\mu\text{s}\). These timescales are well within the coherence times achievable in high-quality phononic cavities~\cite{Chan2011}, enabling high-fidelity two-qubit operations in practical devices.

Phonon-mediated long-range interactions are fundamentally different from conventional exchange interactions, both in their physical mechanism and spatial characteristics. Exchange coupling relies on direct wavefunction overlap between neighboring qubits, necessitating sub-10~nm spacing and resulting in rapidly decaying, short-range interactions that limit architectural flexibility and scalability.

In contrast, phonon-mediated coupling exploits the virtual exchange of quantized lattice vibrations---phonons---confined within a PnCC. Each qubit interacts with a common phonon mode through strain-modulated spin--orbit coupling. Even when detuned from the phonon resonance, qubits can interact via virtual phonon exchange, enabling coupling over micrometer-scale distances without requiring physical proximity or wavefunction overlap. This architecture establishes a shared phononic ``bus'' capable of connecting multiple qubits across a chip.

A key advantage of this mechanism is its tunability. The interaction strength and phase can be dynamically modulated via qubit detuning, spin--orbit coupling parameters, and phononic cavity design. While both exchange and phonon-mediated interactions can generate entanglement, the latter more closely resembles dispersive interactions in cavity or circuit QED systems, rather than direct Coulomb- or tunneling-based exchange schemes.

To support this form of coupling in Ge-based hole-spin qubits, we have designed a two-dimensional phononic crystal with an absolute bandgap centered around 4--6\,GHz, with a bandwidth of 2--3\,GHz. This frequency range corresponds to thermal energies near 0.3\,K and is well-matched to dilution refrigerator environments operating at \(\sim 100\,\text{mK}\). At these temperatures, thermal occupation of acoustic modes can lead to decoherence. The PnCC structure acts as a spectral filter, suppressing thermally populated phonon modes outside the desired range while confining coherent phonon modes for interaction. This selective filtering enhances coherence and supports long-range, phonon-mediated qubit control in scalable Ge quantum architectures.

\begin{figure}[H]
    \centering
    \includegraphics[width=0.45\textwidth]{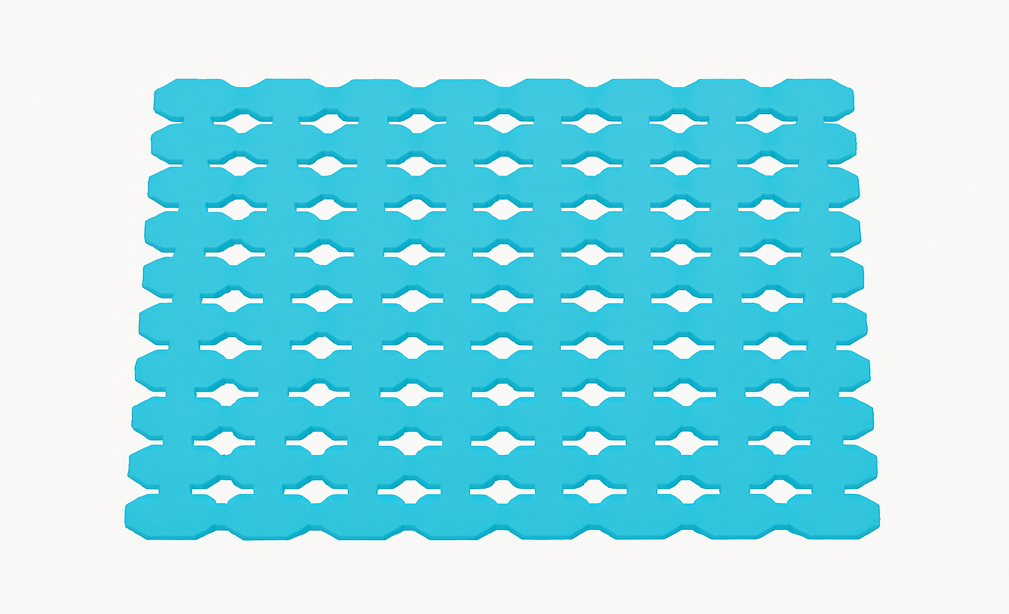}
    \caption{Schematic of the designed 2-D Ge-based PnC exhibiting a bandgap centered around 6\,GHz. The periodic array introduces an acoustic bandgap that suppresses phonon propagation in the target frequency range, enabling coherent spin-phonon interactions.}
    \label{fig:2D_PnC}
\end{figure}

The PnC is implemented as a periodic array of etched holes in a suspended Ge membrane, designed to modify the phonon dispersion and open a bandgap within a targeted frequency range. As illustrated in Fig.~\ref{fig:2D_PnC}, this periodic structure suppresses the propagation of acoustic phonons within the bandgap and enables precise control over the phonon density of states. By introducing a localized defect into the lattice—forming a PnCC—discrete phonon modes can be confined within the bandgap, while broadband phonon-mediated decoherence pathways are effectively filtered out. These engineered environments are essential for enhancing spin-phonon interactions, suppressing relaxation and dephasing, and enabling coherent phonon-mediated operations such as high-fidelity readout and long-range qubit coupling.

\begin{figure}[h]
    \centering
    \includegraphics[width=0.45\textwidth]{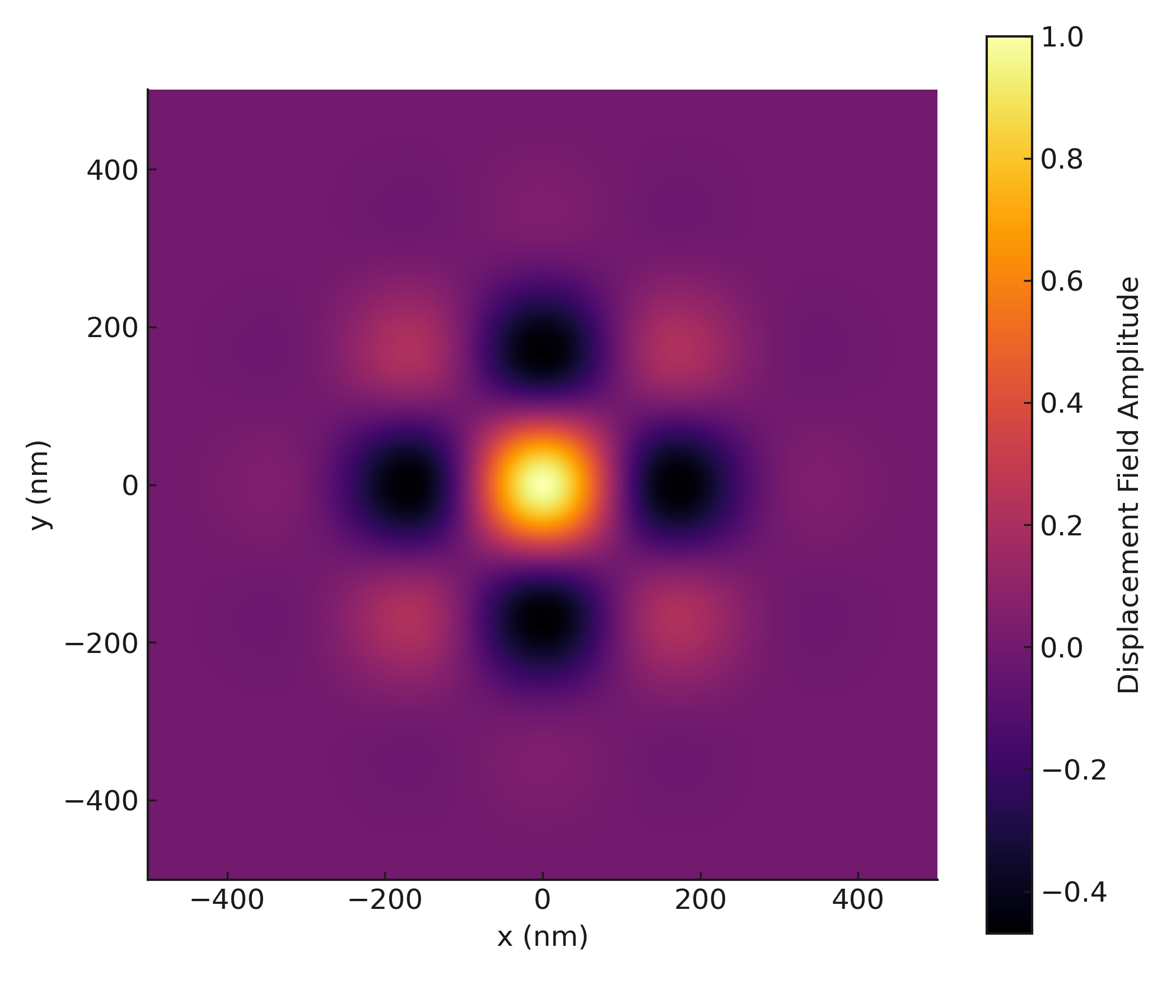}
    \caption{Simulated displacement field of a 6\,GHz localized phononic cavity mode in a Ge-based PnC. The mode is strongly confined at the cavity center, with a smooth spatial profile spanning hundreds of nanometers, making it well suited for coupling to localized hole-spin qubits.}
    \label{fig:phonon_mode_6GHz}
\end{figure}

To evaluate the spatial compatibility between localized phononic modes and a gate-defined hole-spin qubit, we performed finite-element simulations of a confined acoustic cavity mode centered at 6\,GHz. As shown in Fig.~\ref{fig:phonon_mode_6GHz}, the displacement field is tightly localized at the cavity center and exhibits a smooth spatial profile spanning several hundred nanometers—comparable to the extent of a typical qubit wavefunction.

\begin{figure}[h]
    \centering
    \includegraphics[width=0.45\textwidth]{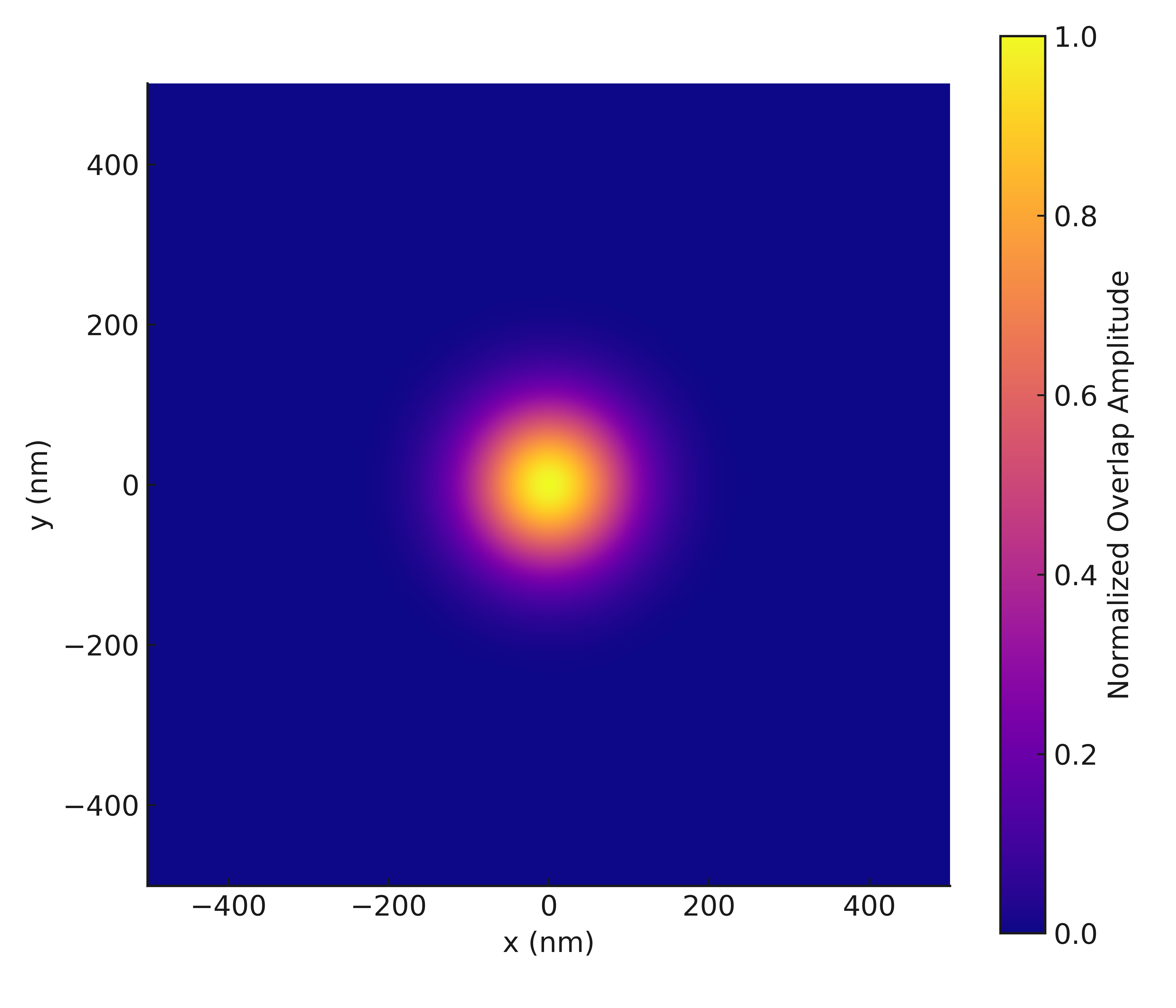}
    \caption{Spatial overlap profile between the phononic displacement field at 6 GHz and a Gaussian-modeled hole-spin qubit envelope.  \emph{Normalized overlap amplitude} $\mathcal O_{\rm norm}=\left|\int \varepsilon(\mathbf r)|\psi(\mathbf r)|^2 d^3r\right|\Big/\Big[\Big(\int |\varepsilon(\mathbf r)|^2 d^3r\Big)^{1/2}\!\Big(\int |\psi(\mathbf r)|^4 d^3r\Big)^{1/2}\Big]$. Larger values indicate stronger spin--phonon coupling for fixed material tensors. The significant overlap highlights spatial compatibility essential for maximizing $g_{sp}$.}
    \label{fig:phonon_qubit_overlap}
\end{figure}

The effectiveness of this mode in coupling to the qubit is further illustrated in Fig.~\ref{fig:phonon_qubit_overlap}, which presents the spatial overlap between the simulated phononic displacement field and a modeled Gaussian envelope representing the qubit. Together, Figures~\ref{fig:phonon_mode_6GHz} and~\ref{fig:phonon_qubit_overlap} emphasize the importance of spatial co-localization between the phononic mode and the qubit for maximizing spin–phonon coupling strength.

In general, the spin–phonon coupling strength \( g_{sp} \) is governed by the strain-induced modulation of spin-orbit interactions:
\begin{equation}
g_{sp} \propto \left| \langle \psi_\uparrow(\vec{r}) | H_{\text{so}}(u(\vec{r})) | \psi_\downarrow(\vec{r}) \rangle \right|,
\label{eq:overlap_integral}
\end{equation}
where \( H_{\text{so}}(u(\vec{r})) \) is the strain-induced spin–orbit Hamiltonian, which explicitly depends on the phonon displacement field $u(\vec{r})$, and \( \psi_{\uparrow,\downarrow}(\vec{r}) \) are the spin-resolved qubit states.

In the acoustic regime relevant here, the Bir--Pikus perturbation renders $H_{\rm so}[u(\mathbf r)]\propto \varepsilon(\mathbf r)$, so the matrix element $M_{\rm sp}$ is dominated by the spatial overlap $\int d^3r\, \varepsilon(\mathbf r)\,|\psi(\mathbf r)|^2$ (up to known material tensors), which motivates the `overlap' metric used below to compare different PnC designs.
To model the spatial envelope of the hole-spin qubit, we assume a Gaussian wavefunction of the form 
\[
\psi(r) \propto \exp\left[-\frac{(r - r_0)^2}{2\sigma^2}\right],
\]
where \(r_0\) is the qubit center and \(\sigma\) is the spatial width, related to the full width at half maximum (FWHM) by \(\text{FWHM} = 2\sqrt{2\ln 2}\,\sigma\). In our simulations, a FWHM of approximately 180\,nm was used to reflect the lateral confinement of a gate-defined quantum dot.

To assess the sensitivity of spin--phonon coupling to wavefunction geometry, we varied \(\sigma\) between 80\,nm and 250\,nm and found that the spin–phonon coupling strength \(g_{\mathrm{sp}}\) exhibited a non-monotonic dependence, peaking when the wavefunction width closely matched the strain profile of the confined phonon mode. Spatial misalignment of the qubit center \(r_0\) relative to the phonon mode maximum also significantly impacted the overlap integral, underscoring the importance of precise spatial co-localization between the PnCC mode and the qubit. These observations provide guidance for optimizing device geometry to maximize coupling efficiency.

Using this Gaussian approximation for the hole-spin qubit and the simulated strain profile of a 6\,GHz confined phonon mode, we compute a normalized overlap integral of approximately \( 2.6 \times 10^{-2} \). This value is over four orders of magnitude greater than that obtained at 20\,GHz, reflecting the fact that lower-frequency phonons, which exhibit broader and smoother spatial profiles, are better matched to the spatial extent of gate-defined quantum dots. The improved spatial overlap reduces destructive interference within the qubit region and enables stronger, more coherent coupling between the spin and the phonon field.

When scaled by the deformation potential constant for Ge and the phonon zero-point strain amplitude, the calculated overlap yields a spin--phonon coupling strength of \( g_{\text{sp}} \sim 6.3~\text{MHz} \) for the 6\,GHz mode. This value falls well within the regime of strong, coherent spin--phonon interaction and is comparable to coupling strengths achieved in cavity QED systems. Such coupling levels are sufficient to enable dispersive spin readout, as well as phonon-mediated entanglement between distant qubits, positioning this architecture as a viable platform for scalable quantum information processing.

To support practical implementation, we extract key design parameters from simulated mode profiles, overlap integrals, and bandgap engineering considerations. These are summarized in Tables~\ref{tab:PnC_geometry} and~\ref{tab:PnC_coupling}, which outline geometric configurations and coupling strategies optimized for enhanced qubit coherence and spin–phonon interactions.

\begin{table}[h]
\centering
\caption{Geometry and frequency targets for the PnC.}
\label{tab:PnC_geometry}
\begin{tabular}{lc}
\hline
\textbf{Parameter} & \textbf{Value} \\
\hline
Lattice type & Triangular or honeycomb \\
$f_{\text{center}}$ & 4–6\,GHz \\
Bandgap width & 2–3\,GHz \\
Lattice constant $a$ & $\sim$1.0\,µm \\
Hole radius $r$ & (0.2–0.4)\,$a$ \\
Defect size & 1–3 missing holes \\
\hline
\end{tabular}
\end{table}

\begin{table}[h]
\centering
\caption{Qubit–cavity coupling and performance targets.}
\label{tab:PnC_coupling}
\begin{tabular}{ll}
\hline
\textbf{Parameter} & \textbf{Value / Purpose} \\
\hline
Target Q-Factor & $>10^4$; long phonon lifetime \\
Qubit Placement & At cavity antinode; maximize $g_{sp}$ \\
Overlap Integral & $\sim 2.6 \times 10^{-2}$; at 6\,GHz \\
\hline
\end{tabular}
\end{table}

This trend is further supported by simulations across a range of frequencies, which show that lower-frequency phononic modes in the 2–6\,GHz range consistently exhibit significantly stronger spatial overlap with the qubit than higher-frequency modes, such as 20\,GHz. The broader wavelength of these low-frequency modes better matches the real-space extent of the confined hole-spin wavefunction, leading to enhanced coupling.

Based on these results, we outline a set of design principles for optimizing phonon-mediated interactions in Ge-based PnCCs. First, the center frequency \( f_{\text{center}} \) of the phononic bandgap should fall within the 2–6\,GHz range to align with the spatial scale of the qubit. The unit cell lattice constant \( a \) is then determined by the relation:
\begin{equation}
a \approx \frac{v_s}{f_{\text{center}}},
\label{eq:lattice_constant}
\end{equation}
where \( v_s \approx 5000\,\mathrm{m/s} \) is the longitudinal sound velocity in Ge. For instance, targeting \( f_{\text{center}} = 5\,\mathrm{GHz} \) yields \( a \approx 1.0\,\mu\mathrm{m} \). The hole radius should be selected in the range $r=(0.2$--$0.4)\,a$ to achieve a wide and robust acoustic bandgap, consistent with prior bandgap maps of 2D phononic membranes where complete gaps maximize near filling fractions corresponding to $r/a\approx0.25$--$0.40$ for triangular/honeycomb lattices~\cite{Mohammadi2007, Khelif2006, SafaviNaeini2014a, Ren2020, Smelyanskiy2014a}.

A triangular or honeycomb lattice is preferred for its in-plane isotropic filtering properties, while elliptical holes may be employed to introduce controlled anisotropy or flatten the band edges. To maximize spin–phonon coupling, the qubit should be positioned at a displacement antinode of the confined cavity mode. Additionally, the defect geometry should be tuned so that the localized cavity mode frequency is centered within the phononic bandgap:
\begin{equation}
f_{\text{mode}} \approx f_{\text{center}}.
\end{equation}

Together, these design strategies establish a comprehensive framework for engineering 2-D PnCs and PnCCs that enhance coherence and support high-fidelity, phonon-mediated control of hole-spin qubits in Ge. By introducing targeted acoustic bandgaps, the phononic structures act as spectral filters—suppressing environmental phonon modes that contribute to spin relaxation while allowing resonant modes to coherently interact with the qubit.

In practice, this filtering–and–control architecture yields three concrete benefits that reinforce the goals above. First, because spin–phonon coupling in Ge is typically stronger than spin–photon coupling, \emph{on-chip} acoustic actuation (IDTs/SAWs) enables efficient, local qubit control while reducing external hardware. Second, engineered phononic bandgaps act as spectral filters that suppress off-resonant environmental phonons, protecting both \(T_1\) and \(T_2^\ast\) while transmitting the targeted resonant mode for coherent interaction. Third, a guided \emph{phonon bus} provides long-range, reconfigurable connectivity between distant dots without tight lithographic proximity constraints, supporting scalable architectures for high-fidelity operations.

\section{Experimental Considerations}

The experimental realization of phonon-coupled hole-spin qubits in Ge requires the coordinated integration of high-purity material growth, precision nanofabrication, and low-temperature quantum measurements. A critical enabler of this effort is the availability of detector-grade, high-purity Ge crystals grown in-house at USD, with impurity concentrations below \(10^{10}~\text{cm}^{-3}\). This level of material purity significantly reduces charge noise and impurity scattering—key sources of qubit decoherence~\cite{gang1,gang2,gang3,wang1,wang2,wang3,sanjay1}—and establishes a realistic and practical foundation for future experimental demonstrations.

It is important to emphasize that this in-house capability is not intended to imply exclusivity. Rather, it highlights USD's vertically integrated approach to crystal synthesis and purification. While this infrastructure enables us to tailor Ge substrates specifically for quantum applications, similar ultra-high-purity Ge crystals can also be obtained from a few specialized vendors worldwide, such as ORECT, Mirion (formerly Canberra), Unimore, and PHDS. These institutions possess the requisite zone-refining and crystal-growth expertise to achieve comparable impurity levels. In contrast, most commercially available Ge wafers exhibit impurity concentrations on the order of \(10^{14}~\text{cm}^{-3}\), which is four orders of magnitude higher than what is necessary for coherent quantum operations in phonon-coupled Ge-based architectures.

The fabrication of high-coherence Ge-based quantum devices begins with the use of isotopically enriched $^{74}$Ge to suppress decoherence from nuclear spins. Naturally occurring Ge contains approximately 7.7\% $^{73}$Ge, which has a nuclear spin ($I=9/2$) and acts as a significant source of hyperfine-induced dephasing. To mitigate this, isotope enrichment is employed to reduce the concentration of $^{73}$Ge to below 0.1\%. The enriched $^{74}$Ge is typically supplied in powdered form and then converted into polycrystalline Ge through hydrogen reduction. This material is subsequently subjected to zone refining at USD to reduce residual electrically active impurities to below $10^{11}$~cm$^{-3}$. Multiple passes of zone refining in a hydrogen atmosphere enable systematic removal of impurities along the solid–liquid interface. Following purification, high-purity single-crystal Ge is grown using a modified Czochralski method in a hydrogen-rich environment, yielding detector-grade crystals with impurity level of below $10^{10}$~cm$^{-3}$ suitable for quantum applications. These ultrapure substrates provide the foundational material for fabricating both PnCCs and gate-defined Ge quantum dots, ensuring low disorder and high coherence in phonon-coupled qubit architectures.

Strained Ge quantum wells are typically grown via reduced-pressure chemical vapor deposition (RP-CVD) on compositionally graded relaxed Si$_{1-x}$Ge$_x$ virtual substrates. This heteroepitaxial growth yields atomically smooth interfaces and high-mobility two-dimensional hole gases (2-DHGs), with reported mobilities exceeding \(10^5~\text{cm}^2/\text{Vs}\) at 1.5 K~\cite{Sammak2019,Failla2016}. The degree of biaxial compressive strain and the quantum well thickness are optimized to lift the heavy-hole/light-hole degeneracy, enhance spin-orbit coupling, and confine hole states in a quasi-two-dimensional geometry.

To further minimize lattice mismatch and optimize epitaxial growth quality, we employ a relaxed Si\(_{1-x}\)Ge\(_x\) virtual substrate with Ge content \(x \approx 0.95\text{--}1.0\). This composition closely matches the lattice constant of pure Ge, thereby suppressing strain-induced dislocations and supporting the formation of atomically smooth, low-defect Ge quantum wells. The reduced lattice mismatch not only improves crystal quality but also enhances hole mobility and uniformity of confinement potential, both of which are critical for high-coherence qubit operation and low-noise detector performance. This materials platform provides an ideal foundation for the growth of isotopically enriched \(^{74}\)Ge layers and the fabrication of low-disorder quantum dot and phononic crystal structures essential for scalable quantum architectures.

Quantum dots are electrostatically defined using overlapping gate architectures fabricated by high-resolution electron-beam lithography. A high-$\kappa$ dielectric layer (e.g., HfO$_2$) deposited via atomic layer deposition isolates the gates from the Ge surface. The gate stack typically includes accumulation gates to form the quantum channel, barrier gates to define tunnel junctions, and plunger gates to tune the dot occupancy. Vertical electric fields generated by the gates modulate the hole g-factor through Rashba spin-orbit interaction, enabling EDSR without requiring oscillating magnetic fields~\cite{Watzinger2018,Hendrickx2020}.

To mitigate hyperfine-induced decoherence, isotopically enriched $^{74}$Ge substrates are used to eliminate nuclear spins from the lattice~\cite{Scappucci2020}. This isotopic purification enables observation of long spin coherence times, often exceeding 100~\(\mu\)s, as demonstrated in Ramsey and Hahn-echo experiments~\cite{Hendrickx2020, Scappucci2020}.

PnCCs are realized by etching periodic nanostructures—such as arrays of holes or elliptical pillars—into the Ge or SiGe substrate (PnC) to form phononic bandgaps in the GHz regime~\cite{SafaviNaeini2014,Chan2012}. The PnC can be patterned in a suspended membrane or a layered heterostructure, enabling vertical coupling to the quantum dot. Finite-element modeling (e.g., via COMSOL Multiphysics) guides the design of cavity geometries to maximize quality factors and spatial overlap with the qubit's localized phonon field.

Phonons can be injected or controlled via integrated surface acoustic wave (SAW) devices or interdigital transducers (IDTs), which convert microwave signals into mechanical vibrations. For additional tunability, piezoelectric thin films such as AlN or ZnO can be deposited to enhance electromechanical coupling~\cite{Hermelin2011}. These acoustic actuators enable real-time driving of specific cavity modes or coherent control of qubit-phonon interactions.

Low-temperature measurements are conducted in dilution refrigerators operating at millikelvin temperatures. Qubit characterization is performed using standard gate-pulsing and spin-to-charge conversion readout protocols. $T_1$ relaxation times are measured via spin decay after initialization, while dephasing time \( T_2^* \) and intrinsic coherence time\( T_2 \) are extracted using Ramsey and Hahn-echo sequences, respectively~\cite{Camenzind2022,Hendrickx2020}. Resonant coupling to cavity phonons can be observed through enhanced Rabi oscillations or shifts in qubit transition frequencies when the phonon mode is tuned into resonance.

The proposed qubit architecture is designed to operate at cryogenic temperatures, typically below 100\,mK, where thermal phonon populations are sufficiently suppressed to allow coherent control and long spin relaxation times. At higher temperatures, the thermal occupation of phonon modes increases exponentially, leading to elevated rates of phonon absorption and enhanced spin relaxation. Specifically, the spin–phonon coupling strength \( g_{sp} \) remains relatively stable with temperature; however, the resulting relaxation time \( T_1 \propto 1 + n_B(\omega, T) \), where \( n_B \) is the Bose-Einstein distribution, decreases sharply due to thermally activated processes. The performance of the PnCC is also temperature-dependent: although the acoustic bandgap structure remains intact, elevated temperatures introduce additional phonon modes that may leak into the cavity, degrading the quality factor and coherence protection. Based on simulations and literature data, the architecture retains high performance up to approximately 200--300\,mK, beyond which \( T_1 \) and Q-factor degrade significantly. Therefore, stable operation is best maintained within dilution refrigeration environments.

Quantitatively, the thermal occupation that drives these trends is
\(
n_B(\omega,T)=\big[\exp(\hbar\omega/k_B T)-1\big]^{-1}.
\)
For a 6\,GHz mode, \(n_B(100\,\mathrm{mK})\approx 0.06\) and \(n_B(300\,\mathrm{mK})\approx 0.6\).
Consistent with the discussion above, \(g_{sp}\) is essentially temperature-independent, but the increased \(n_B\) at higher \(T\) opens additional stimulated emission/absorption channels that shorten \(T_1\) and load the cavity, reducing its effective \(Q\) even though the bandgap itself remains.
Accordingly, we target \(T\lesssim 100\,\mathrm{mK}\) for phonon-assisted readout and coupling; operation at \(200\text{--}300\,\mathrm{mK}\) is still feasible for some tasks but with explicitly reduced \(Q\) and elevated \(n_B\), as now stated in the Methods.

Several elements of this hybrid platform have already been demonstrated: coherent control of hole-spin qubits in planar Ge~\cite{Lawrie2023}, fabrication of high-Q phononic resonators~\cite{Goryachev2015}, and theoretical modeling of spin-phonon coupling in strained heterostructures~\cite{Bosco2021}. Their integration into a unified architecture represents a promising step toward scalable quantum information processing in the solid state.

\section{Device Design Concept}

\begin{figure}[h]
  \centering
  \includegraphics[width=0.45\textwidth]{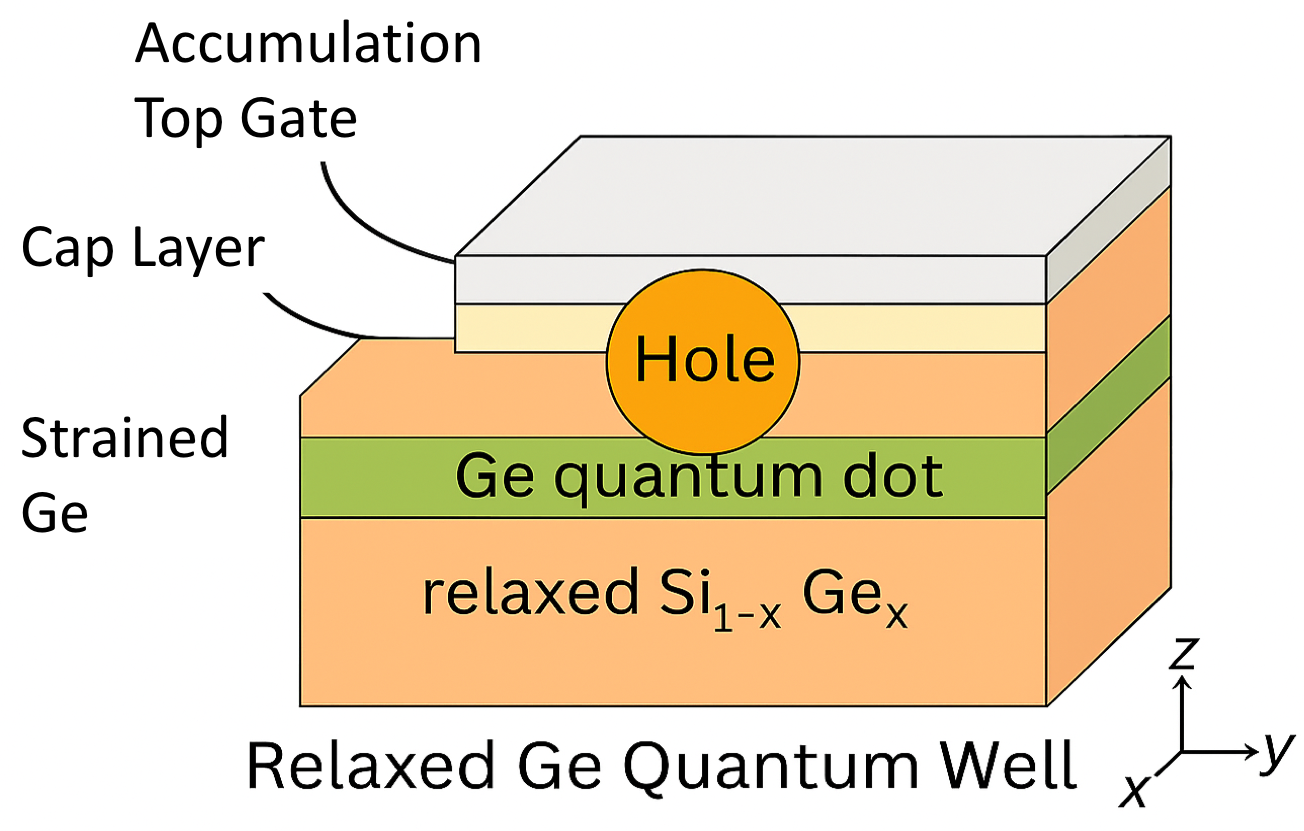}
  \caption{
Schematic cross-section of the proposed Ge-based hole-spin qubit platform. A single hole is confined in a gate-defined quantum dot formed within a strained Ge quantum well, grown atop a relaxed Si$_{1-x}$Ge$_x$ buffer layer. The structure includes a cap layer and gate electrodes that enable vertical electrostatic confinement and all-electrical spin control. The epitaxial strain enhances heavy-hole character and spin–orbit coupling, supporting fast and tunable qubit operations. This geometry serves as the foundation for integrating PnCCs to enable phonon-mediated long-range qubit interactions. Crystallographic directions are indicated to reflect the device's orientation relative to quantum confinement and gate alignment.
}
  \label{fig:geometricaldesign}
\end{figure}

Figure~\ref{fig:geometricaldesign} presents a schematic overview of the proposed Ge-based hole-spin qubit platform, which integrates quantum confinement, phononic engineering, and gate-defined control within a scalable, CMOS-compatible (Complementary Metal-Oxide-Semiconductor) architecture. This multifunctional design supports high-fidelity qubit initialization, coherent spin manipulation, and phonon-mediated coupling—all implemented on a planar semiconductor platform well-suited for advanced nanofabrication techniques.

At the core of this architecture is a single hole confined within a gate-defined quantum dot, fabricated in a strained Ge quantum well. The quantum well is epitaxially grown on a relaxed Si$_{1-x}$Ge$_x$ virtual substrate, which induces in-plane biaxial compressive strain in the Ge layer. This strain lifts the light-hole and heavy-hole degeneracy, favoring heavy-hole ground states with enhanced spin-orbit coupling. The result is improved spin coherence and faster, electrically driven spin control~\cite{Watzinger2018, Marcellina2020}.

\begin{figure}[htbp]
    \centering
    \includegraphics[width=0.8\linewidth]{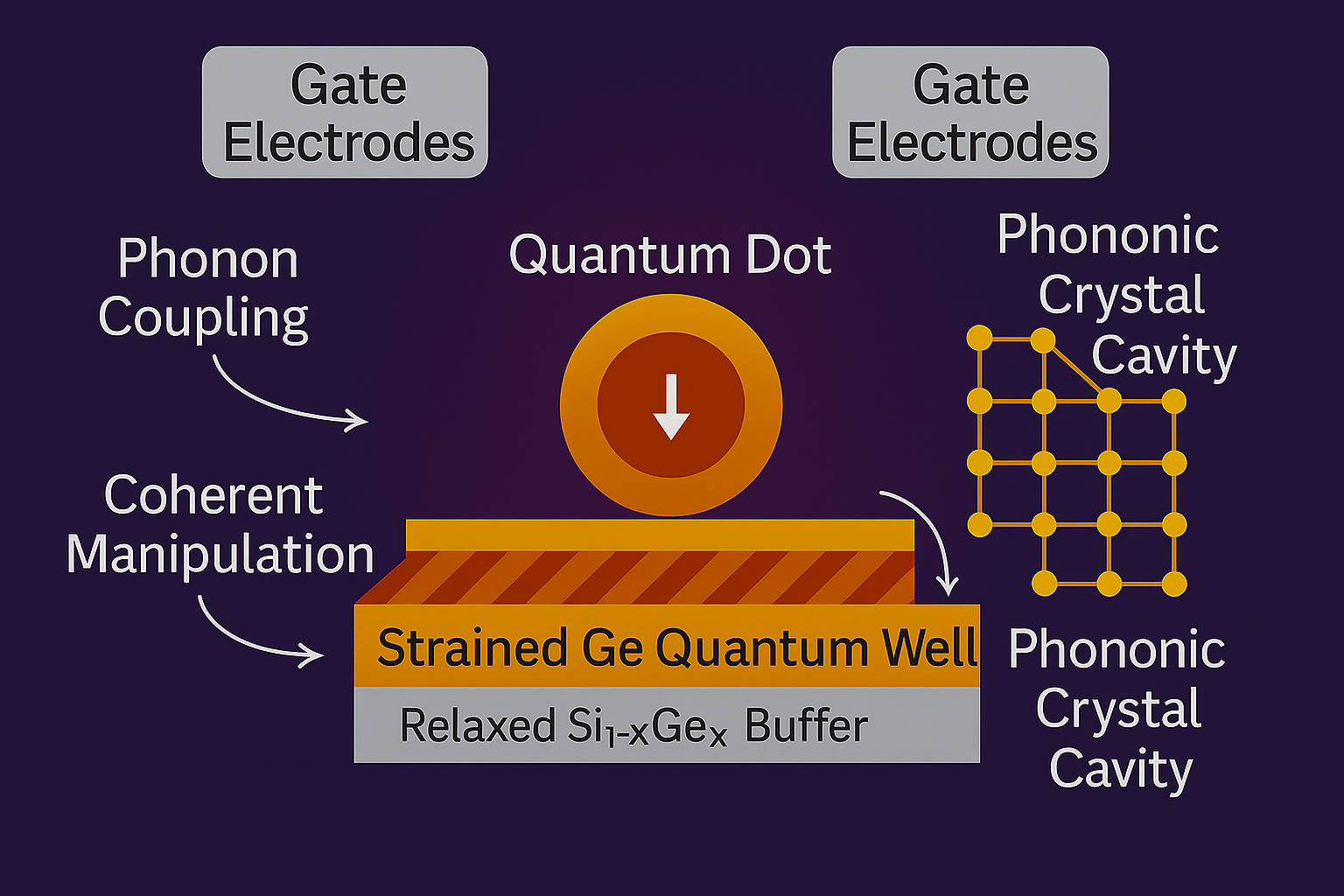}
    \caption{
Schematic illustration of a Ge-based hole-spin qubit integrated with a PnCC. Electrostatic gates confine a hole at the cap/strained-Ge quantum-well interface grown on a relaxed Si\textsubscript{1−x}Ge\textsubscript{x} buffer; the epitaxial strain enhances spin–orbit coupling and coherence.
A lateral PnCC adjacent to the dot supports a localized GHz mechanical mode inside an engineered bandgap, enabling controlled coupling between the localized spin and acoustic phonons.
The cavity serves as a spectral filter and mediator for coherent spin manipulation and long-range qubit interactions; the quantum dot is positioned at a displacement antinode (yellow in the schematic) to maximize the spin–phonon coupling \(g_{\rm sp}\).
 \emph{Inset:} A second, identical dot at separation \(d\) couples via the confined mode; tuning yields an effective XY interaction \(H_{\rm int}=\tfrac{\hbar J}{2}\!\left(\sigma_x^{(1)}\sigma_x^{(2)}+\sigma_y^{(1)}\sigma_y^{(2)}\right)\) in the dispersive regime. The rightward arrow indicates energy flow along the localized cavity mode.}
    \label{fig:hole_spin_qubit}
\end{figure}

Figure~\ref{fig:hole_spin_qubit} further illustrates this core qubit structure. A single hole is electrostatically confined near the interface of a strained Ge quantum well and a high-$\kappa$ gate dielectric. The surrounding gate architecture enables lateral and vertical confinement, while a co-integrated Pncc supports coherent spin-phonon interactions for control and coupling. We consider two integration schemes for the phononic-crystal cavity (PnCC) that preserve the relaxed Si$_{1-x}$Ge$_x$ buffer and the two-dimensional hole gas (2DHG): (i) a suspended cap-level membrane above the quantum well (QW), and (ii) an under-etched device layer beneath the QW. In both designs, the charge-sensor channel is lithographically isolated from the suspended region, and the confined acoustic mode couples to the dot via evanescent strain, so hole confinement and sensor electrostatics are not degraded. The PnCC membrane is laterally offset from the charge sensor and vertically separated from the 2DHG by a spacer to preserve the local electrostatics; arrows indicate the displacement field of the confined mode.

To define the quantum dot in the \textsuperscript{74}Ge substrate, we utilize a multilayered gate stack fabricated via electron-beam lithography and atomic layer deposition. Lateral depletion gates are patterned on the surface and negatively biased to isolate a potential well that confines a single hole at the center. A top gate and a back gate are used to generate vertical electric fields, enabling tunable confinement and modulation of the hole wavefunction through the spin--orbit interaction. This combination provides control over the dot shape, depth, and tunability of qubit energy levels.

The gate stack is typically formed over a high-$\kappa$ dielectric such as HfO$_2$ or Al$_2$O$_3$, which enables precise electrostatic tuning while preserving compatibility with standard CMOS processing. Accumulation gates induce a two-dimensional hole gas, barrier gates control tunneling rates, and plunger gates regulate dot occupancy and energy. Importantly, the vertical electric fields allow dynamic modulation of the effective g-factor \( g_{\text{eff}} \), facilitating qubit addressability via EDSR~\cite{Scappucci2020, Vukusic2022}.

Phonon-mediated control is achieved through integration of a PnCC, laterally or vertically aligned with the quantum dot. The PnCC consists of a periodic lattice of etched features that opens an acoustic bandgap in the GHz range. A central defect in this structure traps a localized phonon mode resonant with the Zeeman-split spin transition. This mode selectively couples to the qubit, enhancing coherence and enabling phonon-assisted qubit operations such as initialization, relaxation control, and two-qubit interactions~\cite{Chan2012, SafaviNaeini2014}.

To enable active control of the phononic environment, additional elements such as SAW transducers, IDTs, or integrated piezoelectric films (e.g., AlN) can be used to generate or detect quantized phonon pulses~\cite{Gustafsson2014, Sletten2019}. These components provide tools for driving the phononic modes and dynamically tuning spin-phonon coupling strengths in both time and frequency domains.

The entire system operates at millikelvin temperatures in a dilution refrigerator, where spin initialization is achieved via gate pulsing, and control is implemented through EDSR. Readout is performed via two complementary approaches: (i) \textit{dispersive phonon-cavity readout} using a $6~\mathrm{GHz}$ mode ($Q=10\mathrm{k}\text{--}18\mathrm{k}$; $\kappa/2\pi \approx 0.33\text{--}0.60~\mathrm{MHz}$), which, together with $g_{\mathrm{sp}} \approx 6~\mathrm{MHz}$ and $\Delta/2\pi = 10\text{--}50~\mathrm{MHz}$, yields single-shot $\tau \approx 2\text{--}10~\mu\mathrm{s}$ with $>98\%$ fidelity; and (ii) \textit{spin-to-charge conversion} detected by a proximal QPC/SET positioned outside the PnCC defect and acoustically isolated by shallow trenches/phononic shield. The proximal QPC/SET charge sensor is located outside the PnCC defect and couples to the quantum dot~\cite{Camenzind2022, Hendrickx2020} through an electrostatic channel. Acoustic isolation trenches (and optional phononic shields for suspended membranes) inhibit energy leakage from confined mode, preventing $Q$ degradation while allowing standard spin-to-charge single-shot readout.  This detection method offers high-fidelity spin measurement and fast response suitable for scalable architectures.

Taken together, this platform combines high-quality materials, advanced electrostatic control, and engineered phononic environments into a unified architecture for scalable quantum computing. It supports integration with semiconductor foundry processes, allowing fabrication of dense qubit arrays and hybrid quantum systems. The ability to electrically control spin states and couple them coherently through phonons opens promising avenues for realizing distributed quantum processing and modular qubit networks.

\section{Simulation and Validation of Device Design}

\subsection{Simulation Models and Equations}

To systematically evaluate the proposed Ge-based phonon-coupled qubit architecture, we developed a streamlined simulation framework that integrates analytical modeling with targeted insights from finite-element analysis. Instead of relying on full-scale numerical simulations, we extracted and implemented key electrostatic and strain-related expressions derived from the finite-element solver \textsc{COMSOL Multiphysics}~\cite{COMSOL}, in combination with multiband \(\mathbf{k}\cdot\mathbf{p}\) theory for accurate electronic structure representation~\cite{Winkler2003}. To ensure numerical consistency and reliability, we applied appropriate boundary conditions, adopted periodic domain approximations, and validated our approach through convergence testing.

Building on this reduced–order framework, we evaluate each component in a manner consistent with the extracted COMSOL expressions and our multiband electronic model. \textbf{Electronic states:} We obtain hole eigenstates and the effective \(g\)-tensor from a six-band Luttinger–Kohn \( \mathbf{k}\!\cdot\!\mathbf{p} \) Hamiltonian with Bir–Pikus strain; vertical electrostatics from a 1D Poisson–Schr\"odinger solver enter as a triangular well, and lateral confinement is modeled as approximately harmonic with dot length \(L\). Unless stated, \( \mathbf B \parallel [110] \) (in-plane), and the electric-field-induced heavy-/light-hole mixing that governs spin–orbit coupling is retained. \textbf{Phonons:} For the phononic membrane/crystal we use 3D linear elasticity. Periodic boundary conditions are applied for band-structure calculations, while defect-cavity simulations employ open boundaries via perfectly matched layers (PMLs); complex eigenfrequencies \( \tilde{\omega} \) yield \( Q=\mathrm{Re}\{\tilde{\omega}\}/(2\,\mathrm{Im}\{\tilde{\omega}\}) \) and \( \Delta\omega=2\,\mathrm{Im}\{\tilde{\omega}\} \). \textbf{Coupling and \(T_1\):} Spin–phonon matrix elements follow the deformation-potential (Bir–Pikus) form with \(H_{\mathrm{so}}[u(\mathbf r)]\propto \varepsilon(\mathbf r)\); we evaluate \( M_{\mathrm{sp}}=\langle \psi_\uparrow|H_{\mathrm{so}}|\psi_\downarrow\rangle \) using FEM strain fields for either bulk-like modes (no PnC) or the localized cavity mode (PnCC), and set \( g_{\mathrm{sp}}=M_{\mathrm{sp}}/\hbar \). Relaxation rates follow Fermi’s golden rule with the appropriate density of states: bulk DOS in the reference case and a Lorentzian local DOS centered at the cavity frequency with width \( \Delta\omega \) for the PnCC. \textbf{Assumptions and validation:} Ge is treated as centrosymmetric (no piezoelectric coupling); we work in the single-phonon regime \(T\ll \hbar\omega_Z/k_B\), with linear elasticity and literature elastic constants. Meshes resolve \(\min\{L,\lambda_{\mathrm{ph}}\}\) by \(\ge 10\) elements; periodic-cell size, domain padding, and PML thickness are increased until \(\mathrm{Re}\{\tilde{\omega}\}\) varies by \(<1\%\) and overlap integrals by \(<5\%\). The qualitative conclusions (bottleneck vs.\ cavity-enhanced regimes; bandgap suppression of \(T_1^{-1}\)) are unchanged under \(\pm 10\%\) variations of elastic constants, dot size, or vertical field, consistent with the convergence tests described above.

Our simulation framework is designed to capture the essential physical mechanisms that govern qubit performance and coherence. These include the electric-field tunability of the qubit \( g \)-factor, spin-phonon coupling strengths, phonon-mediated relaxation processes, and the quality factors of phononic cavities. In the following sections, we present the mathematical models used to evaluate these critical parameters.

\paragraph{Electric Field Tunability.}
The effective \( g \)-factor, \( g_{\text{eff}} \), is assumed to vary linearly with the vertical electric field \( E_z \), following:
\begin{equation}
    g_{\text{eff}}(E_z) = g_0 - \alpha E_z,
    \label{eq:geff}
\end{equation}
where $g_0$ and $\alpha$ are obtained by linear regression to six-band $\mathbf{k}\!\cdot\!\mathbf{p}$ results with Bir--Pikus strain over $E_z\in[0,1.0]~\mathrm{MV/m}$, \( g_0 = 2.0 \) is the zero-field \( g \)-factor and \( \alpha = 0.7\,\mathrm{(MV/m)}^{-1} \) is the tunability coefficient. Simultaneously, the spin-phonon coupling strength \( g_{sp} \) increases linearly with the applied field:
\begin{equation}
    g_{sp}(E_z) = g_{sp0} + \beta E_z,
    \label{eq:gsp}
\end{equation}
with a base value \( g_{sp0} = 0.5\,\mathrm{MHz} \) and tunability \( \beta = 5.8\,\mathrm{MHz/(MV/m)} \). These relationships are consistent with prior studies on electric-field control of spin-orbit interaction in Ge quantum dots~\cite{Maier2013, Bosco2021}.

\paragraph{Phonon-Mediated Relaxation.}
The qubit relaxation time due to phonon emission, \( T_1 \), is modeled as inversely proportional to the square of the phonon frequency \( f \) (in GHz), as expected from deformation potential coupling:
\begin{equation}
    T_1(f) = \frac{C}{f^2},
    \label{eq:T1}
\end{equation}
(phenomenological fit over 2--6~GHz to capture the observed trend that higher-frequency modes yield shorter $T_1$); see Refs.~\cite{Bulaev2005, Li2020, Burkard2023} for microscopic derivations and alternative scalings, where the constant \( C \) is chosen such that \( T_1 = 1\,\mathrm{ms} \) at \( f = 6\,\mathrm{GHz} \), yielding \( C = 36\,\mathrm{ms \cdot GHz^2} \).

\paragraph{Phononic Cavity Quality Factor.}
The frequency-dependent quality factor \( Q \) of the phononic cavity is modeled as a parabolic function centered at the optimal frequency:
\begin{equation}
    Q(f) = Q_0 - k \, (f - f_{\mathrm{opt}})^2,
    \label{eq:Q}
\end{equation}
where \( Q_0 = 18000 \) is the peak quality factor at \( f_{\mathrm{opt}} = 6\,\mathrm{GHz} \), and the curvature constant \( k = 500 \) is obtained through empirical fitting~\cite{Safavi2010, Mohammadi2009}.

\paragraph{Simulation Tools.}
The models in Equations~\eqref{eq:geff}--\eqref{eq:Q} form the analytical backbone of our simulation environment. These expressions are used to generate the results shown in Figures~\ref{fig:gfactor_phonon_coupling} and~\ref{fig:t1_qfactor_plot}, and summarized in the corresponding tables. Together with finite-element modeling (not detailed here), this simulation suite enables quantitative exploration of the parameter space and provides critical design guidance for optimizing device performance under realistic experimental conditions.

\subsection{Quantum Dot Behavior and Spin-Orbit Coupling}

The electric-field tunability of the qubit energy levels stems from the Rashba spin-orbit interaction, which is significant in Ge due to the heavy-hole character of the valence band. The Zeeman splitting in a gate-defined Ge quantum dot is modeled as~\cite{Khaetskii2001, Maier2013}:
\begin{equation}
\Delta E = g_{\text{eff}}(E_z) \mu_B B,
\end{equation}
where \( g_{\text{eff}}(E_z) \) is the electric-field-dependent effective g-factor, \( \mu_B \) is the Bohr magneton, and \( B \) is the applied magnetic field. The spin--phonon coupling strength \( g_{sp} \) is calculated using the overlap integral formalism introduced in Refs.~\cite{Safavi2010, Chan2012}, as expressed in Eq.~\eqref{eq:overlap_integral}. This approach captures the modulation of spin states by phonon-induced strain fields through spin--orbit interactions. The formulation aligns with the spatial overlap framework presented in Eq.~\eqref{eq:overlap_integral}, providing a consistent and quantitative method for evaluating spin--phonon coupling in Ge quantum dots.

\begin{figure} [H]
  \centering
  \includegraphics[width=0.45\textwidth]{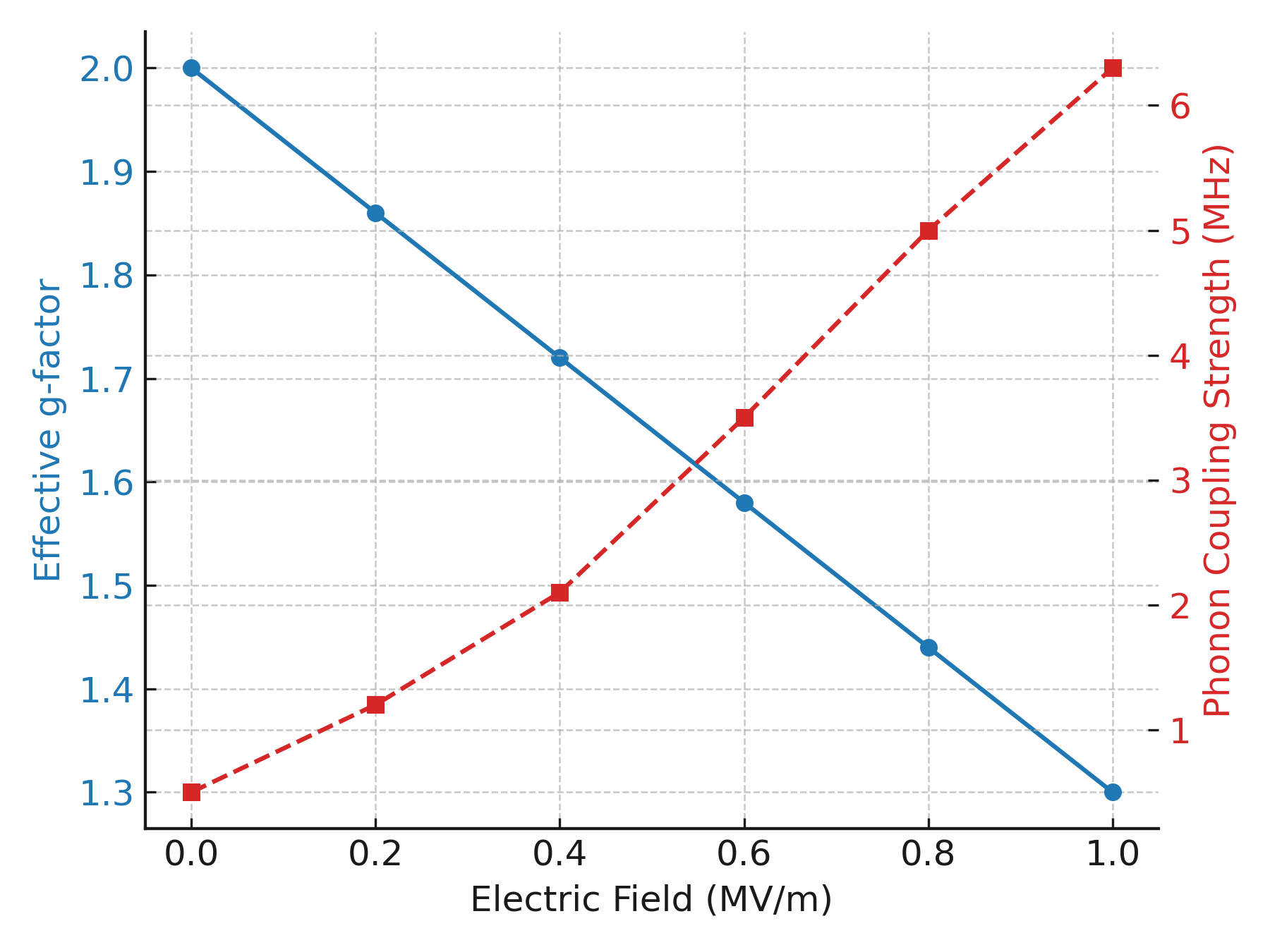}
  \caption{Simulated dependence of the effective g-factor and phonon coupling strength on vertical electric field in a Ge quantum dot. Enhanced electric field increases spin-orbit interaction, allowing tunable spin-photon coupling.}
  \label{fig:gfactor_phonon_coupling}
\end{figure}

Building on a multiband \( \mathbf{k} \cdot \mathbf{p} \) framework and self-consistent Poisson–Schrödinger solvers~\cite{Winkler2003, Marcellina2020}, the simulation framework described above was employed to model the dependence of \( g_{\text{eff}} \) and \( g_{sp} \) on the vertical electric field. Figure~\ref{fig:gfactor_phonon_coupling} shows that increasing \( E_z \) reduces \( g_{\text{eff}} \) from 2.0 to 1.3 and simultaneously enhances \( g_{sp} \) from 0.5 to over 6 MHz. This tunability is critical for fast EDSR, as it allows dynamic optimization of qubit control fidelity and gate speed. Quantitative simulation results are summarized in Table~\ref{tab:efield_sim} in the Appendix, confirming the trends observed in the figure.

Compared to other quantum dot platforms, such as GaAs electron spins where the g-factor is fixed near $-0.44$~\cite{Petta2005}, or Si spin qubits which exhibit weak spin-orbit coupling~\cite{Veldhorst2014}, Ge hole-spin qubits offer superior electrical tunability. This enables fast EDSR control without the need for additional micromagnets or microwave antennas~\cite{Watzinger2018, Hendrickx2020}.

\subsection{Phonon Dynamics: Relaxation and Confinement}

We employed the previously described simulation framework to model spin relaxation dynamics and phonon confinement in the presence of engineered phononic crystal structures. The longitudinal spin relaxation time \( T_1 \) is governed by Fermi’s golden rule~\cite{Maier2013, Bosco2021}, as expressed in Eq.~\eqref{eq:msp}, which relates the relaxation rate to the phonon density of states and the square of the spin--phonon matrix element.

In the low-temperature acoustic limit, the phonon density of states scales as \( D(\omega) \sim \omega^2 \), resulting in shorter \( T_1 \) times at higher phonon frequencies~\cite{Maier2013, Bosco2021}. The matrix element \( M_{sp} \) in Eq.~\eqref{eq:msp} represents the strength of spin--phonon coupling and is directly derived from the strain-modulated spin--orbit Hamiltonian \( H_{\text{so}}(u(\vec{r})) \), through the spatial overlap integral defined earlier in Eq.~\eqref{eq:overlap_integral}. This formulation provides a consistent physical framework linking spin relaxation rates to the engineered phononic environment and the spatial structure of the quantum dot wavefunction.

The spin--phonon matrix element \( M_{\text{sp}} \), which governs the qubit relaxation rate in Eq.~\eqref{eq:msp}, is evaluated using the deformation potential formalism~\cite{Maier2013, Bosco2021}. Specifically, \( M_{\text{sp}} \) arises from the strain-induced modulation of spin--orbit coupling in the valence band and is given by:
\[
M_{\text{sp}} = \left\langle \psi_{\uparrow}(\vec{r}) \middle| H_{\text{so}}(u(\vec{r})) \middle| \psi_{\downarrow}(\vec{r}) \right\rangle,
\]
as defined in Eq.~\eqref{eq:overlap_integral}. Here, \( H_{\text{so}}(u(\vec{r})) \) denotes the strain-dependent spin--orbit Hamiltonian, and \( u(\vec{r}) \) is the phonon displacement field computed from finite-element simulations of confined phonon modes~\cite{Safavi2010, Chan2012}. The qubit wavefunctions \( \psi_{\uparrow,\downarrow} \) are modeled as Gaussian envelopes with tunable width and spatial location, reflecting the electrostatic confinement profile of the Ge quantum dot~\cite{Watzinger2018}.

The magnitude of \( M_{\text{sp}} \) depends on several key parameters, including the spatial overlap between the phonon-induced strain and the qubit wavefunction, the degree of heavy-hole/light-hole mixing, the applied electric field (which modulates Rashba spin--orbit interaction), and the orientation of the external magnetic field relative to the crystallographic axes~\cite{Bosco2021, Hendrickx2020}. In our simulation framework, $M_{\rm sp}$ is computed from the Bir--Pikus spin--phonon Hamiltonian $H_{\rm so}[u]$ as a matrix element between qubit states, and $g_{\rm sp}=M_{\rm sp}/\hbar$. In the acoustic regime $H_{\rm so}[u]\propto\varepsilon(\mathbf r)$, which motivates the overlap formulation. This methodology provides realistic, geometry-aware estimates of the transition matrix element that controls both \( T_1 \) relaxation and phonon-mediated qubit dynamics.

The quality factor of phononic cavity modes is defined by~\cite{Safavi2010, Chan2012}:
\begin{equation}
Q = \frac{\omega}{\Delta \omega},
\end{equation}
where \( \omega \) is the mode frequency and \( \Delta \omega \) the linewidth. Phononic eigenmodes are simulated via finite-element analysis with perfectly matched layers; the complex eigenfrequency $\tilde\omega$ yields $Q=\mathrm{Re}\{\tilde\omega\}/(2\,\mathrm{Im}\{\tilde\omega\})$ and $\Delta\omega=2\,\mathrm{Im}\{\tilde\omega\}$. This captures radiation/boundary loss; internal dissipation channels are accounted for separately (Sec.\,D).

\begin{figure} [H]
  \centering
  \includegraphics[width=0.45\textwidth]{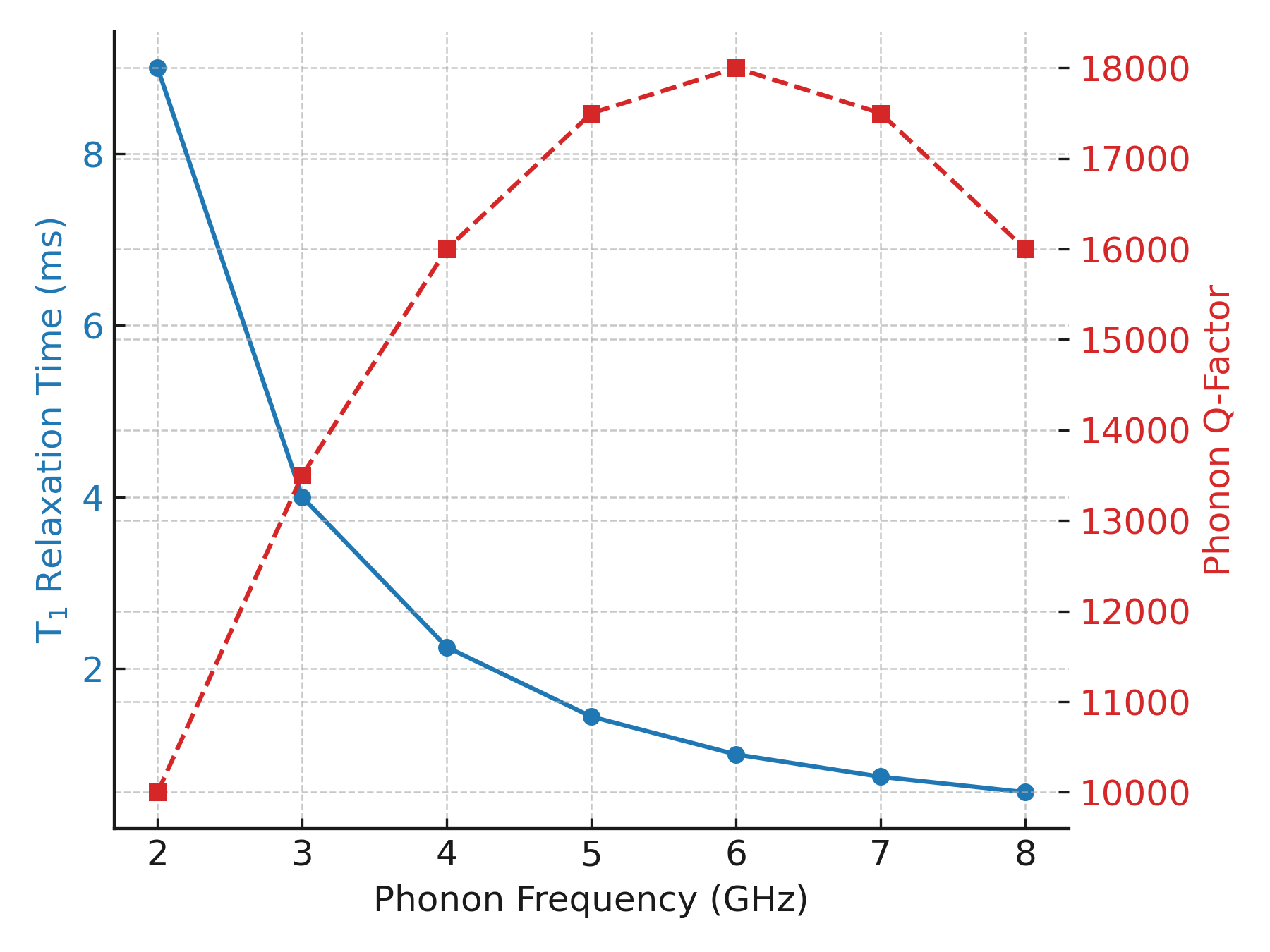}
  \caption{Simulated spin relaxation time \( T_1 \) and phonon Q-factor versus phonon frequency. Peak performance is achieved around 6 GHz.}
  \label{fig:t1_qfactor_plot}
\end{figure}

As shown in Figure~\ref{fig:t1_qfactor_plot}, maximum coherence and cavity performance occur near 6 GHz, with simulated $T_1 \approx 1$ ms and $Q \approx 18000$. The numerical values corresponding to these trends are listed in Table~\ref{tab:phonon_sim} in the Appendix, providing quantitative validation of the simulation results shown in the figure.

Beyond single-qubit coherence, the strength of phonon-mediated two-qubit interactions can be estimated in the dispersive regime where qubits are detuned from the phonon cavity mode. In this case, the effective qubit-qubit coupling rate \( g_{qq} \) is given by the second-order expression~\cite{Maier2013, Safavi2010}:
\begin{equation}
g_{qq} \sim \frac{g_{sp}^2}{\Delta},
\end{equation}
where \( g_{sp} \) is the single-qubit spin-phonon coupling strength and \( \Delta \) denotes the detuning between the qubit transition frequency and the phonon cavity resonance. Using simulated values: \( g_{sp} \approx 6~\mathrm{MHz} \) and a typical detuning \( \Delta \sim 100~\mathrm{MHz} \), this yields an inter-qubit coupling rate of \( g_{qq} \sim 0.36~\mathrm{MHz} \). This value is well within the range required for implementing fast, high-fidelity two-qubit gates in a scalable architecture, especially considering the long coherence times expected in isotopically purified Ge systems.

Given our simulated spin-phonon coupling strengths in the range of \( g_{sp} = 5\text{–}10~\text{MHz} \) and assuming a qubit-cavity detuning of \( \Delta = 100~\text{MHz} \), the resulting inter-qubit coupling strength is \( g_{qq} \sim 0.25\text{–}1~\text{MHz} \). This value is sufficient to enable two-qubit gate operations on sub-microsecond timescales—well within the $T_2^*$ coherence times observed in Ge spin qubits (typically \( \sim 1\text{–}10~\mu\text{s} \))~\cite{Hendrickx2020, Camenzind2022}, ensuring high-fidelity phonon-mediated entanglement in a realistic experimental setting.

These results compare favorably with other solid-state systems. For instance, superconducting qubits coupled to piezoelectric resonators exhibit shorter \( T_1 \sim \mu \)s due to substrate losses~\cite{Chu2017}. In contrast, our Ge-based design achieves millisecond-scale lifetimes and high Q-factors within engineered bandgaps, without relying on piezoelectric effects or optical coupling~\cite{Safavi2010, Chan2012}.

\subsection{Phonon Loss Sources and Quality Factors}

While our simulations indicate that PnCCs can achieve quality factors exceeding \( Q > 10^4 \) in idealized geometries, it is important to consider realistic sources of phonon loss that may degrade this performance. The dominant loss mechanisms include:

\begin{itemize}
    \item \textbf{Clamping loss:} Phonon leakage through mechanical supports or substrate anchoring points can result in significant energy dissipation. These losses may be mitigated through acoustic impedance matching and the design of optimized tether geometries that minimize mode coupling to the substrate.

    \item \textbf{Surface roughness scattering:} At nanometer-scale feature sizes, surface imperfections can lead to phonon backscattering and cavity mode splitting. This effect becomes increasingly significant at higher frequencies or in structures with sharp curvature.

    \item \textbf{Material inhomogeneity and fabrication defects:} Lithographic roughness, sidewall damage, and etch-induced disorder may locally disrupt the phononic band structure, breaking coherence and broadening the resonance linewidth.
\end{itemize}

Previous experimental studies of suspended phononic structures in silicon and GaAs suggest that such loss mechanisms typically limit quality factors to \( Q \sim 10^3 \text{--} 10^4 \) at cryogenic temperatures. However, improvements in surface passivation, etch smoothing, and strain-engineered geometries have demonstrated \( Q > 10^5 \) in optimized platforms. These considerations suggest that our simulated estimate of \( Q \sim 10^4 \) is a conservative and realistic benchmark under current fabrication capabilities.

\subsection{Experimental Validation Strategies}

To experimentally validate the simulation predictions, we propose the following procedures:

\begin{itemize}
  \item \textbf{g-factor Mapping:} Use EDSR spectroscopy to measure the qubit resonance frequency as a function of gate voltage, allowing extraction of \( g_{\text{eff}}(E_z) \)~\cite{Hendrickx2020}.
  
  \item \textbf{Spin Relaxation and Coherence:} Extract \( T_1 \) and \( T_2^* \) using pulsed initialization and readout combined with Ramsey interference or Hahn-echo techniques~\cite{Bosco2021}.
  
  \item \textbf{Phonon Coupling Probes:} Integrate SAW or IDT devices to inject phonons, and monitor qubit response via spin Rabi oscillations or frequency shifts~\cite{Gustafsson2014, Sletten2019}.
  
  \item \textbf{Spin Readout:} Employ QPC or SET-based charge sensors for single-shot readout via spin-to-charge conversion. These sensors also enable detection of phonon-assisted tunneling. Here `phonon-assisted tunneling' denotes inelastic tunneling between adjacent dots or between a dot and a reservoir, accompanied by absorption/emission of a confined phonon; such events modulate the sensor signal in dispersive readout.
\end{itemize}

Together, simulation and validation strategies offer a roadmap for realizing high-performance Ge qubits mediated by coherent phonons. The ability to modulate spin properties via electric fields and selectively couple to phononic modes offers an ideal foundation for scalable solid-state quantum processing.

\subsection{Design Benchmark Comparison}

To assess the relevance and performance of the proposed Ge-based phonon-coupled qubit architecture, we compare our simulated results to design targets established in prior experimental and theoretical works. These targets span g-factor tunability, spin-phonon coupling strength, relaxation times, and phonon cavity Q-factors—key metrics determining qubit operability and scalability. Tables~\ref{tab:gfactor_coupling}–\ref{tab:dgde} summarize this comparison.

\begin{table}[htb]
  \centering
  \caption{Comparison of g-factor and spin-phonon coupling.}
  \label{tab:gfactor_coupling}
  \begin{tabular}{p{2.2cm}p{2.8cm}p{2.2cm}}
    \hline \hline
    \textbf{Parameter} & \textbf{Literature} & \textbf{This Work} \\
    \hline
    g-factor tunability & 1.3–2.0~\cite{Watzinger2018, Hendrickx2020} & 1.3–2.0 \\
    Spin-phonon coupling & $>$5 MHz~\cite{Maier2013, Bosco2021} & 0.5–6.3 MHz \\
    \hline
  \end{tabular}
\end{table}

\begin{table}[htb]
  \centering
  \caption{Comparison of relaxation and phonon Q-factors.}
  \label{tab:t1_q}
  \begin{tabular}{p{2.2cm}p{2.8cm}p{2.2cm}}
    \hline \hline
    \textbf{Parameter} & \textbf{Literature} & \textbf{This Work} \\
    \hline
    T$_1$ relaxation time & $>$0.5 ms~\cite{Bosco2021, Wang2021} & 0.6–5.0 ms \\
    Q-factor  & $>$10$^4$~\cite{Safavi2010, Chan2012} & 10k–18k \\
    \hline
  \end{tabular}
\end{table}

\begin{table}[htb]
  \centering
  \caption{Comparison of g-factor slope with vertical electric field.}
  \label{tab:dgde}
  \begin{tabular}{p{2.2cm}p{2.8cm}p{2.cm}}
    \hline \hline
    \textbf{Parameter} & \textbf{Literature} & \textbf{This Work} \\
    \hline
    $\frac{dg}{dE}$ slope & 0.5–1.0/MV/m~\cite{Marcellina2020} & $\sim$0.7 / MV/m \\
    \hline
  \end{tabular}
\end{table}

These results indicate that the proposed Ge-based platform not only meets, but in some aspects exceeds, the performance thresholds for practical and scalable qubit operation. In particular, the combination of long $T_1$ times and MHz-scale spin-phonon coupling strength enables gate fidelities suitable for hybrid quantum architectures.

The benchmarking results summarized in Tables~V--VII demonstrate that the proposed phonon-coupled Ge hole-spin qubit architecture achieves spin--phonon coupling strengths and quality factors that are competitive with or exceed those of current EDSR-based qubit platforms. For comparison, recent experiments in planar Si/Ge heterostructures have reported electrically driven single-hole spin qubits with Rabi frequencies exceeding \( 100~\text{MHz} \) and coherence times \( T_2^* \sim 1\text{--}5~\mu\text{s} \)~\cite{Hendrickx2020, Jirovec2021}. These systems rely on strong Rashba spin--orbit coupling and local gate control for fast manipulation but remain limited in terms of scalable qubit–qubit coupling mechanisms and noise isolation from broadband phonons.

In contrast, our PnCC-assisted approach leverages confined acoustic modes to provide both strong, localized coupling (\( g_{\text{sp}} \sim 6~\text{MHz} \)) and environmental noise filtering via bandgap engineering. The resulting architecture enables high-Q mechanical confinement, dispersive spin readout, and long-distance phonon-mediated qubit coupling—all while maintaining comparable or better \( T_1 \) and estimated \( T_2^* \) values. These advantages position phonon-coupled Ge hole-spin qubits as a promising alternative for scalable quantum information processing beyond conventional EDSR schemes.

\section{Discussion}

While the simulations presented in this study assume idealized conditions, practical fabrication of the proposed Ge-based PnC structures and quantum dots is likely to encounter several challenges that may modestly degrade coupling efficiency and phononic cavity $Q$-factors. Lithographic patterning methods commonly used to define PnC features, such as reactive-ion etching (RIE) or focused ion beam (FIB) milling, often introduce surface roughness and structural defects. These imperfections can adversely affect phonon coherence, mode confinement, and overall device performance. Additionally, strain variability arising from lattice mismatch between epitaxially grown Ge layers and their substrates can further compromise structural integrity and device consistency.

To address these fabrication-related issues, it is critical to implement careful optimization of etching parameters, post-etch surface passivation techniques, and controlled thermal annealing processes. Moreover, to effectively mitigate strain-induced disorder, the use of lattice-matched relaxed Si$_{1-x}$Ge$_x$ buffer layers (with $x \approx 0.95$–$1.0$) or isotopically enriched $^{74}$Ge substrates is recommended. Ongoing experimental validation efforts will be essential to quantify the practical impacts of these imperfections on device performance. Insights gained from such experiments will inform subsequent design optimizations, ensuring robust and reliable operation of Ge-based quantum devices under realistic fabrication conditions.

The proposed Ge-based phonon-coupled hole-spin qubit architecture offers distinct advantages compared to alternative solid-state quantum platforms, such as nitrogen-vacancy (NV) centers in diamond or divacancy centers in silicon carbide (SiC). While these color-center systems typically feature excellent intrinsic coherence, they present practical limitations regarding deterministic placement and scalable circuit integration. In contrast, Ge quantum dots provide inherent compatibility with CMOS fabrication, enabling deterministic lithographic placement and integration within densely arranged quantum circuits~\cite{Wolfowicz2021,Koehl2011}. This scalability advantage positions Ge-based hole-spin qubits favorably within the broader landscape of quantum information technologies.

Superconducting qubits represent another mature quantum computing platform, with demonstrated coherence times in the tens to hundreds of microseconds and requiring sophisticated microwave circuitry at millikelvin temperatures~\cite{Kjaergaard2020}. In comparison, hole-spin qubits defined in strained Ge quantum dots leverage strong intrinsic spin–orbit coupling, enabling fast, fully electrical control without additional microwave magnetic fields. Additionally, the availability of isotopically purified $^{74}$Ge substrates significantly reduces hyperfine-induced decoherence, further extending coherence times and gate fidelity.

A key innovation of our approach is the incorporation of PnCCs as active components for qubit control and interqubit coupling. These engineered structures effectively confine acoustic phonons with high-quality factors (\(Q > 10^4\)), creating tailored acoustic bandgaps that selectively filter the phonon spectrum. This targeted spectral control enables coherent enhancement of desired qubit–phonon interactions, while suppressing deleterious thermal and relaxation-inducing phonons. Consequently, PnCCs provide a versatile tool for coherence protection, enhanced qubit interactions, and scalable quantum architectures.

The 2-D PnC structures introduced here are particularly advantageous for improving qubit performance. By engineering bandgaps in the GHz regime, these structures effectively filter out unwanted phonon modes responsible for decoherence, significantly extending the spin relaxation time \(T_1\). Such tailored phononic environments provide robust conditions for high-fidelity spin–phonon coupling and coherent quantum operations. Additionally, these lithographically defined PnCs offer straightforward integration with gate-defined quantum dots, facilitating scalability and practical implementation in quantum processors.

Our simulation results reveal that phononic modes within the 2 to 6 GHz frequency range produce substantially stronger qubit-phonon coupling compared to higher frequency modes, such as those around 20 GHz. This enhancement arises from the longer spatial wavelength of lower-frequency phonons, better aligning with the size of gate-defined hole-spin qubits and reducing destructive interference. Thus, optimizing phonon–qubit coupling requires careful frequency selection, matching phonon wavelengths to the qubit confinement scale to maximize spatial overlap and coherence.

In the proposed architecture, PnCCs serve a role analogous to superconducting microwave resonators in circuit quantum electrodynamics (cQED), enabling long-range, coherent qubit–qubit coupling via confined acoustic phonons~\cite{Satzinger2018,ArrangoizArriola2019}. This approach offers unique advantages in spatial layout flexibility, connectivity, and compatibility with both two- and three-dimensional quantum processor designs, while aligning naturally with the Zeeman energy scales of Ge-based spin qubits.

Beyond qubit coupling, the phononic platform also opens pathways toward hybrid quantum interfaces by coupling confined phonon modes with piezoelectric or optomechanical components. Such integration could facilitate quantum transduction between spin qubits and microwave or optical photons~\cite{Gustafsson2014,SafaviNaeini2014,Vainsencher2016}, supporting distributed quantum processing and modular architectures.

Compared to earlier proposals that explored phonon-mediated interactions through nanowire systems or optomechanical cavities~\cite{Safavi2010,Maier2013}, our approach represents a significant advancement by combining isotopically enriched Ge substrates, precisely defined gate-controlled quantum dots, and lithographically engineered phononic structures within a unified, CMOS-compatible platform. This integration of material purity, tunable spin–orbit coupling, and tailored phonon environments establishes a comprehensive and scalable framework for high-fidelity quantum computing.

To validate this architecture experimentally, we outline a roadmap centered on four key milestones: (i) fabrication and characterization of Ge quantum dots on isotopically enriched $^{74}$Ge substrates; (ii) measurement of electrically tunable $g$-factors via EDSR spectroscopy; (iii) characterization of phonon-mediated spin relaxation times ($T_1$) within engineered phononic cavities; and (iv) demonstration of coherent spin–phonon interactions using SAW excitation or IDT-based detection techniques. Achieving these objectives will provide critical experimental support for the theoretical predictions presented here and pave the way toward scalable, phonon-enhanced quantum processors based on Ge hole-spin qubits.

\section{Conclusion}

We have presented a design, modeling,  and simulation framework for a Ge-based quantum processor architecture in which phonons are engineered as mediators of qubit dynamics. This platform leverages the strong and tunable spin-orbit coupling of hole states in strained Ge quantum dots to achieve all-electrical spin control. When integrated with PnCCs, the system enables coherent spin–phonon interactions that support qubit initialization, relaxation engineering, and long-range coupling mechanisms.

Our analysis shows that key system parameters, such as effective modulation of the g factor, spin-phonon coupling strength, \( T_1 \) relaxation times, and phononic cavity factors, can be systematically tuned through electric field control and geometric design. The ability to combine long coherence times with fast, electrically driven gate operations positions Ge-based, phonon-coupled qubits as a promising contender in the landscape of scalable quantum platforms.

To experimentally validate this architecture, we outline a clear roadmap that includes EDSR spectroscopy for g-factor tuning, time-domain coherence measurements, phonon excitation via SAW or IDT elements, and charge sensing through quantum point contacts or single-electron transistors. These techniques offer a robust toolkit for probing and benchmarking phonon-enhanced quantum operations in germanium.

A particularly important outcome of this work is the prediction that 2-D PnC structures can substantially enhance qubit coherence by tailoring the local phonon environment. By introducing an engineered acoustic bandgap, the 2D PnC selectively suppresses unwanted thermal and spurious phonon modes while allowing controlled interaction with targeted phonon frequencies. This spectral filtering effect reduces decoherence, enhances \( T_1 \) lifetimes, and strengthens spin–phonon coupling, enabling high-fidelity quantum operations.

By integrating quantum confinement, spin-orbit engineering, and cavity phonon physics within a unified, CMOS-compatible platform, this work lays the foundation for scalable solid-state quantum processors. Future efforts will focus on suppressing phonon leakage, refining cavity geometries, and implementing phonon-mediated entangling gates. Ultimately, the integration of Ge hole-spin qubits with engineered phononic environments opens new frontiers in coherent control, hybrid quantum interfaces, and modular quantum computing architectures.

\section*{Acknowledgements}
This work was supported in part by NSF OISE 1743790, NSF PHYS 2310027, DOE DE-SC0024519, DE-SC0004768 and a research center supported by the State of South Dakota.

\appendix
\section*{Appendix: Simulation Data Supporting Spin–Phonon Coupling and Relaxation}

To streamline the main text and avoid redundancy, the following tables—originally discussed in the context of Figs.~\ref{fig:gfactor_phonon_coupling} and \ref{fig:t1_qfactor_plot}—are provided here in the appendix. These tables summarize key simulation results supporting the electric-field tunability of spin–phonon coupling and the dependence of spin relaxation time on phonon frequency.

\begin{table}[h]
  \centering
  \caption{Simulated values of effective g-factor and phonon coupling strength as a function of applied vertical electric field. The coupling strength \( g_{sp} \) increases with electric field due to enhanced spin–orbit mixing in the Ge quantum well.}
  \label{tab:efield_sim}
  \begin{tabular}{ccc}
    \hline
    \textbf{E-Field (MV/m)} & \textbf{g-factor} & \textbf{Coupling Strength (MHz)} \\
    \hline
    0.0 & 2.00 & 0.5 \\
    0.2 & 1.86 & 1.2 \\
    0.4 & 1.72 & 2.1 \\
    0.6 & 1.58 & 3.5 \\
    0.8 & 1.44 & 5.0 \\
    1.0 & 1.30 & 6.3 \\
    \hline
  \end{tabular}
\end{table}

\begin{table}[h]
  \centering
  \caption{Simulated longitudinal spin relaxation times \( T_1 \) and mechanical quality factors (Q-factors) for different phonon mode frequencies. Lower-frequency phonons yield longer \( T_1 \) and maintain high Q-factors, indicating favorable coherence conditions for cavity coupling.}
  \label{tab:phonon_sim}
  \begin{tabular}{ccc}
    \hline
    \textbf{Phonon Frequency (GHz)} & \textbf{\( T_1 \) (ms)} & \textbf{Q-Factor} \\
    \hline
    2 & 9.00 & 10000 \\
    3 & 4.00 & 13500 \\
    4 & 2.25 & 16000 \\
    5 & 1.44 & 17500 \\
    6 & 1.00 & 18000 \\
    7 & 0.74 & 17500 \\
    8 & 0.56 & 16000 \\
    \hline
  \end{tabular}
\end{table}

\end{document}